\documentclass[onefignum,onetabnum, onealgnum]{siamart190516}

\headers{A Distributed-Memory Algorithm for Heavy-Weight Perfect Matching}{A. Azad, A. Bulu\c{c}, X. Li,  X. Wang, and J. Langguth}

\usepackage{cite}
\usepackage{amsmath,amssymb,amsfonts}

\usepackage[right=1.1in]{geometry}
\usepackage{afterpage}
\usepackage{graphics}
\usepackage{graphicx}
\usepackage{textcomp}
\usepackage{booktabs}
\usepackage{longtable}
\usepackage{color}
\def\BibTeX{{\rm B\kern-.05em{\sc i\kern-.025em b}\kern-.08em
    T\kern-.1667em\lower.7ex\hbox{E}\kern-.125emX}} 

\usepackage{algorithm}
\usepackage[noend]{algpseudocode}

\usepackage{url}

\usepackage{subfig}
\usepackage{amsmath}
\usepackage{amssymb}
\usepackage{multirow}

\usepackage{enumitem}

\newcommand{\W}{\mathcal{W}}

\newcommand{\M}{\mathcal{M}}

\newcommand{\Am}{\mathbf{A}}

\newcommand{\revision}[1]{{\color{black}#1}}
\newcommand{\lastrevision}[1]{{\color{black}#1}}
\newcommand{\finalrevision}[1]{{\color{black}#1}}

\newcommand\iid{i.i.d.}

\title{A distributed-memory algorithm for computing a heavy-weight perfect matching on
bipartite graphs}

\author{Ariful Azad\thanks{Intelligent Systems Engineering, Indiana University Bloomington, IN, USA ({\tt azad@iu.edu}).}
  \and Aydin Bulu\c{c}\thanks{CRD, Lawrence Berkeley National Laboratory, CA, USA ({\tt abuluc@lbl.gov}).}
  \and Xiaoye S. Li\thanks{CRD, Lawrence Berkeley National Laboratory, CA, USA ({\tt xsli@lbl.gov}).}
    \and Xinliang Wang\thanks{Tsinghua University,
Beijing, China ({\tt clarencetc@163.com}). }
  \and Johannes Langguth\thanks{Simula Research Laboratory, Fornebu, Norway ({\tt langguth@simula.no}). }}

\begin{document}
\sloppy
\maketitle

\begin{abstract}
We design and implement an efficient parallel algorithm for finding a perfect matching in a weighted bipartite graph such that weights on the edges of the matching are large.
This problem differs from the maximum weight matching problem, for which scalable approximation algorithms are known. It is primarily motivated by finding good pivots in scalable sparse direct solvers before factorization. Due to the lack of scalable alternatives, distributed solvers use sequential implementations of maximum weight perfect matching algorithms, such as those available in MC64. 
To overcome this limitation, we propose a fully parallel distributed memory algorithm that first generates a perfect matching and then iteratively improves the weight of the perfect matching by searching for weight-increasing cycles of length four in parallel.
For most practical problems the weights of the perfect matchings generated by our algorithm are very close to the optimum.
An efficient implementation of the algorithm scales up to 256 nodes (17,408 cores) on a Cray XC40 supercomputer and can solve instances that are too large to be handled by a single node using the sequential algorithm.
\end{abstract}


\section{Introduction}\label{sec:intro}
The maximum cardinality matching (MCM) problem is a classical topic in combinatorial optimization. Given a graph, it asks for a set of non-incident edges of maximum size.  
For the bipartite version of the problem, efficient sequential algorithms such as  \emph{Hopcroft-Karp}~\cite{Hopcroft} have been known for a long time. 
Practical algorithms for bipartite MCM
have recently been studied intensively~\cite{DuffKayaUcar,LAHM14,tpds16}, leading to the development of scalable distributed-memory algorithms~\cite{LPM2010PC,matchingipdps16}. 

Finding a maximum-cardinality matching on a bipartite graph that also has maximum weight (also called the {\em assignment problem} in the literature) is a harder problem, both 
w.r.t.~complexity and in practice. 
The latter is due to the fact that because 
in the {\em transversal problem} (i.e., maximum cardinality matching), any augmenting path is sufficient  for augmenting the matching. In the assignment problem, the algorithm effectively has to verify that an augmenting path is of minimum length, which might require searching through a significant part of the graph. Thus, only cardinality matchings allow concurrent vertex-disjoint searches. As a result, the assignment problem tends to be much harder to parallelize. Recent published attempts at parallelizing the assignment problem, e.g.~\cite{sathe2012auction} rely on the auction paradigm~\cite{Bertsekas}. While this approach has demonstrated some speedups, its scalability is limited and it is inefficient in maximizing the cardinality in distributed memory~\cite{riedy2010}. 

In this paper, we follow a different approach. Instead of relaxing both the maximum cardinality and maximum weight requirements at the same time, we only relax the maximum weight requirement and use an algorithm that always returns maximum cardinality. Thus, we solve the transversal problem optimally and the assignment problem approximately.  We only consider graphs that have perfect matchings. \revision{Hence, we relax the maximum weight requirement of a {\em maximum-weight perfect matching} (MWPM) and thus solve the {\em heavy-weight perfect matching} (HWPM) problem on distributed memory machines.} 

The motivation for this problem comes from sparse direct solvers. Often, sparse linear systems are too large to be solved in a single node, necessitating distributed-memory solvers such as SuperLU\_DIST~\cite{li2003superlu_dist}. Partial pivoting, which is stable in practice, requires dynamic row exchanges that are too expensive to perform in the distributed case. Consequently, distributed-memory solvers often resort to static pivoting where the input is pre-permuted to have a ``heavy'' diagonal so that the factorization can proceed without further pivoting. The definition of ``heavy'' at the minimum implies having only nonzeros on the diagonal. Whether maximizing the product or the sum of absolute values of the diagonal is the right choice for the objective function is debatable, but both can be solved by finding a perfect bipartite matching of maximum weight. In this formulation, rows and columns of the sparse matrix serve as vertices on each side of the bipartite graph, and nonzeros as edges between them. 

Since the input matrix is already distributed as part of the library requirements, it is necessary to use distributed-memory parallel matching algorithms. 
\revision {However, the lack of scalable matching algorithms forces distributed-memory solvers to assemble the entire instance on a single node and then use a sequential matching library, such as the highly-optimized implementations of MWPM algorithms  available in MC64~\cite{MC64}.
For instance, the new algorithm in SuperLU\_DIST demonstrated strong scaling to $24,000$ cores~\cite{superlu3D}, but still used the sequential static pivoting step.
Such reliance on a sequential library is disruptive to the computation, infeasible for larger instances, and certainly not scalable.}


We use the distributed memory parallel cardinality matching algorithm from our prior work~\cite{matchingipdps16}, and combine it with 
a distributed memory parallel algorithm that improves the weights of perfect matchings. 
Inspired by a sequential algorithm by Pettie and Sanders~\cite{pettieSanders}, our algorithm relies on finding short weight-increasing paths. 
In our case, these paths are cycles of length 4 since we maintain a perfect matching.

In this manner, we get a scalable algorithm for the overall problem 
that can be used in the initialization of sparse direct solvers, although it is not restricted to that application. 

The main contributions of this paper are as follows:
\begin{itemize}
\item{\bf Algorithm:}  We present a highly-parallel algorithm for the heavy weight perfect bipartite matching problem. 
For most practical problems (over 100 sparse matrices from the SuiteSparse matrix collection~\cite{ufget} and other sources), the weights of the perfect matchings generated by HWPM are within 99\% of the optimum solution.

\item {\bf Performance:} We provide a hybrid OpenMP-MPI implementation that runs significantly faster than a sequential implementation of the exact algorithm (up to $2500\times$ faster on 256 nodes of NERSC/Cori).
On 256 nodes of the same system, the parallel implementation attains up to $114\times$ speedup relative to its running time on a single node.
The HWPM code is freely distributed as part of the Combinatorial BLAS library~\cite{combblas_web}.

\item {\bf Impact:} The presented algorithm can be used to find good pivots in distributed sparse direct solvers such as SuperLU\_DIST, eliminating a longstanding performance bottleneck. 
The HWPM code has been successfully integrated with the SuperLU~\cite{superlu_web} and STRUMPACK~\cite{strumpack_web} solvers.

\end{itemize}


\section{Related Work} 
\label{s:related}

The bipartite maximum cardinality matching problem has been studied for more than a century, and many different algorithms for solving it have been published over the years \cite{Hopcroft,ABMP,PF,Goldberg}. Experimental studies \cite{DuffKayaUcar,KLMU2012} established that when using heuristic initialization \cite{LMS2010}, optimized variants of two different approaches, the \emph{Pothen-Fan} algorithm \cite{PF} and the \emph{push-relabel} algorithm \cite{Goldberg} provide superior practical performance in the sequential case.
Both algorithms have efficient shared memory counterparts \cite{azad2012multithreaded,LAHM14} which show good scaling on a single compute node. For distributed memory systems however, the problem has proven to be extremely challenging. Due to the inherent sequentiality of the problem (i.e.,~no theoretically efficient parallel algorithm is known), such parallel algorithms tend to require a large number of consecutive communication rounds. 
More recently, a \emph{push-relabel} variant that exploits the fact that local matching can be performed at a much faster rate than nonlocal operations was presented~\cite{LPM2010PC}. A different approach formulated the problem in terms of sparse matrix operations~\cite{matchingipdps16}. An implementation of the resulting algorithm scaled up to $16,384$ cores. Our work uses this implementation as a subroutine.

For the weighted case, parallel approximation algorithms have been shown to scale very well \cite{suitor}, even in distributed memory \cite{pointer}. Furthermore, these algorithms also work for nonbipartite graphs. On the other hand, exact algorithms such as \emph{successive shortest paths} have proven difficult to parallelize, even for shared memory. Currently, auction algorithms \cite{sathe2012auction,Bertsekas}, which essentially constitute a weighted version of the \emph{push-relabel} algorithm, are a promising direction and can efficiently find matchings of near-maximum weight, but they tend to be very inefficient at finding in distributed memory \cite{riedy2010}, expecially w.r.t.~finding perfect cardinality matchings. In shared memory however, they can be competitive \cite{HFVTP2012}. For that case, Hogg and Scott showed that the auction algorithm provides matrix orderings for direct linear solvers of a quality similar to the exact method~\cite{HoggScott2015}.

The aim of our work is similar to Hogg and Scott's. However, we target solvers with static pivoting such as SuperLU which require a perfect matching. Furthermore, we target distributed memory machines and thus need to develop a different algorithm. Pettie and Sanders described and analyzed sequential linear time $2/3-\epsilon$ approximation algorithms for the general weighted matching problem. 
\revision{Our idea of using cycles of length 4 is inspired by this work.}

\section{Notation and Background} 
\label{s:back}

For any matrix $\Am$ of size $n \times n'$ there is a weighted bipartite graph $G=(R \cup C, E, w)$, whose vertex set consists of $n$ row vertices in $R$ and $n'$ column vertices in $C$. For each nonzero entry $a_{ij}$ of $\Am$, $E$ contains an undirected edge $\{i,j\}$ that is incident to row vertex $i$ and column vertex $j$, and has a weight $w(\{i,j\}) = a_{ij}$.
The weight of a set of $k$ edges $\{e_1, e_2,..., e_k\}$ is simply the sum of weights of individual edges: $w(\{e_1, e_2,..., e_k\}) = \sum_{i=1}^k w(e_i)$.

Given a bipartite graph $G=(R \cup C, E, w)$, a \emph{matching} $\M$ is a subset of $E$ such that at most one edge in $\M$ is incident on any vertex.
Given a matching $\M$ in $G$, an edge is matched if it belongs to $\M$, and unmatched otherwise. Similarly, a vertex is matched if it is an endpoint of a matched edge. If an edge $\{i,j\}$ is matched, we define $m_j:=i$ and $m_i:=j$ and call the vertices \emph{mates} of each other. A matching $\M$ is \emph{maximal} if there is no other matching $\M'$ that properly contains $\M$, and $\M$ is \emph{maximum} if $|\M|{\geq}|\M'|$ for every matching $\M'$ in $G$. Furthermore, if $|\M| = |R| = |C|$, $\M$ is called a {\it perfect} matching. When $\Am$ has full structural rank, the corresponding bipartite graph $G$ has a perfect matching. Since this paper focuses on perfect matching algorithms, we assume $|R| = |C|$. We denote $|R|$ by $n$ and $|E|$ by $m$. When referring to matrices, we use $\mathit{nnz}$ instead of $m$ to denote the number of nonzeros in the matrix.
Now, the {\it perfect matching} problem consists of either finding a matching that is perfect, or deciding that no such matching exists in the input graph.

\revision{
The weight $w(\M)$ of a matching $\M$ is the sum of the weights of its edges. 
The {\it maximum weight matching} problem asks for a matching of maximum weight regardless of cardinality,
while the {\em maximum weight perfect matching} (MWPM) problem asks for a perfect matching that has maximum weight among all perfect matchings. If we multiply all weights by $-1$, the MWPM problem becomes equivalent to  the {\em minimum weight perfect matching} problem~\cite{schrijver2003combinatorial}. For selecting pivots in linear solvers, only the absolute value of the nonzero elements matters. Thus we can assume all weights to be positive.
}


All the above problems can be further subdivided into the bipartite and general case, with the latter often requiring significantly more sophisticated algorithms.
In the following, we will restrict 

\lastrevision{
Given a matching $\M$ in $G$,  a path $P$ is called an \emph{$\M$-alternating path} if the edges of $P$ are alternately matched and unmatched.
Similarly, an \emph{$\M$-alternating cycle} in $G$ is a cycle whose edges are alternately matched and unmatched in $\M$. 
A $k$-cycle is a cycle containing $k$ vertices and $k$ edges.
An $\M$-augmenting path is an \emph{$\M$-alternating path} that connects two unmatched vertices. 
We will simply use alternating and augmenting paths if the associated matching is clear from the context.
Let $\M$ and $P$ be subsets of edges in a graph. 
Then, the symmetric difference between  $\M$ and $P$ is defined by
\begin{equation*}
    \M\oplus P := (\M\setminus P)\cup (P\setminus\M).
\end{equation*}
If $\M$ is a matching and $P$ is an $\M$-augmenting path or an $\M$-alternating cycle, then $\M\oplus P$ is also a matching, and this operation is called \emph{augmentation}.
For these two cases, we say ``$\M$ is augmented by $P$".  
A matching $\M$ can be simultaneously augmented by a set of $k$ vertex-disjoint augmenting paths or alternating cycles  $\{P_1, P_2,..., P_k\}$ as follows:
\begin{equation*}
    \M\oplus \{P_1, P_2,..., P_k\} = \M\oplus P_1 \oplus P_2 \oplus ... \oplus P_k.
\end{equation*}

Augmenting a matching by an augmenting path increases the matching cardinality by one.
By contrast, augmenting a matching $\M$ by an alternating cycle $P$ does not change the cardinality of $\M\oplus P$, but $w(\M\oplus P)$ can be different from $w(\M)$.
An \emph{$\M$-alternating cycle} $P$ is called a \emph{weight-increasing cycle} if $w(\M\oplus P)$ is greater than $w(\M)$.
The gain $g$ of an $\M$-alternating cycle $P$ is defined by
\begin{equation*}
    g(P) := w(\M\oplus P) - w(\M).
\end{equation*}
If the gain $g(P)$ is positive, then $P$ is called a \emph{weight-increasing cycle}. 
In our parallel algorithm, we only consider weight-increasing cycles of length four. 
When an alternating cycle is formed by the vertices $i$ and $j$ and their mates $m_i$ and $m_j$, we use $g(i,j,m_j,m_i)$ to denote the gain of that 4-cycle.
In this case, the gain of the alternating $4$-cycle can be expressed more concisely by
\begin{equation*}
g(i,j,m_j,m_i)=w(\{i,j\})+w(\{m_i,m_j\})-w(\{i,m_i\})-w(\{m_j,j\}).
\end{equation*}

}


Since we study a maximization problem, for any $\alpha \in (0,1)$, we say  
that a perfect matching $\M$ in $G$ is $\alpha$-optimum if $w(\M) \geq \alpha w(\M^*)$, where $\M^*$ is a perfect matching of maximum weight in $G$. We call an algorithm 
an $\alpha$-approximation algorithm if it always returns $\alpha$-optimum solutions.

\section{A Sequential Linear Time Algorithms} 
\label{sec:algoPS}
\lastrevision{
The problem of approximating maximum weight perfect matching is significantly more difficult than approximating only maximum weight matching. Clearly, its complexity is bounded from below by the complexity of finding a perfect matching, which is $O(\sqrt{n}m)$ for the Hopcroft-Karp \cite{Hopcroft} or Micali-Vazirani algorithm \cite{MicaliVazirani1980}. For several special cases of the problem, faster algorithms are known. The complexity for any useful\footnote{Useful in the sense of yielding lower complexity than the exact solution.}~approximation algorithm is also bounded from above since a maximum weight perfect matching can be found in time $O(\sqrt{n}m\log(nN))$ where $N$ is the magnitude of the largest integer edge weight \cite{DuanPettieSu2018}, making the gap between both bounds comparatively small. To the best of our knowledge, no approximation algorithm for the problem with a complexity lower than $O(\sqrt{n}m\log(nN))$ is known today. 

A perfect matching $\M$ is a maximum-weight perfect matching if there is no weight-increasing alternating cycle in the graph~\cite{MC64}. Based on this principle, we can design a simple MWPM algorithm that starts from a perfect matching and improves the weight by augmenting the current matching with weight-increasing alternating cycles. 
For optimality, this algorithm has to search for cycles of all lengths, which can be expensive because a cycle can span over the entire graph in the worst case. By contrast, if we augment a perfect matching only with length-limited cycles (e.g., cycles whose lengths are less than $k$), we can expedite the search, but the final perfect matching may not have the optimum weight. In this paper, we took the latter route, where our algorithm repeatedly augments a perfect matching by weight-increasing alternating cycles of shorter lengths. 
Since our goal is to design a practical algorithm suited for parallelization, we require that the cycle finding process should not be slower than the perfect matching algorithm.
In practice, this means that its typical running time should be no more than twice that of the perfect matching algorithm. Previous work has shown that good practical algorithms usually find perfect matchings
in close to linear time \cite{DuffKayaUcar,KLMU2012}. Thus, for our purposes, it is desirable to use a linear time algorithm for improving the weight. 

A $\frac{2}{3}-\epsilon$ sequential linear time approximation algorithm for the weighted matching problem on general graphs was presented by Pettie and Sanders~\cite{pettieSanders}. It relies on the fact that alternating paths or cycles of positive gain containing at most two matched edges can be found in $O(deg(v))$ time, where $deg(v)$ is the maximum degree among the vertices in the path or cycle. We adapt this technique for perfect matchings by restricting the search to cycles. Such cycles will always be of length $4$.
While it is certainly possible to search for longer cycles, doing so would not result in a linear time algorithm~\cite{pettieSanders}.  

Note that the approximation guarantee of the Pettie-Sanders algorithm for maximum weight matching does not carry over to MWPM. To see this, consider a $6$-cycle with edges that alternate between low and high weight. Clearly, it allows two perfect matchings, one having low and one having an arbitrarily higher weight. It also does not contain any $4$-cycle. Thus, no approximation guarantee can be obtained for an algorithm that only uses $4$-cycles\footnote{The same is true for any cycle of bounded length. However, considering cycles of unbounded length would essentially result in an optimal MWPM algorithm with  higher time complexity}. However, in doing so we obtain the practical sequential Algorithm~\ref{alg:detseq} which will serve as the basis for the parallel algorithm.

\begin{algorithm}[t]
\begin{small}
\caption{The deterministic sequential algorithm adapted by Pettie and Sanders~\cite{pettieSanders}}
\begin{algorithmic}[1]
\State Let $\M$ be a perfect matching
  \While{$iteration \leq maxiters$} 
		\State $S:=\emptyset$
		\ForAll{$j \in C$} 
			\State Find an alternating cycle $(i,j,m_i,m_j)$ with the maximum positive gain $g(i,j,m_i,m_j)$ if one exists
			\State $S$:=$S\cup (i,j,m_i,m_j)$ 
		\EndFor	

        \State Compute a vertex-disjoint set $D(S)$
        \If{$D(S)=\emptyset$} \State {\textbf{break}}
        \EndIf
        \State $\M:=\M \oplus$ $D(S)$
	\EndWhile
\end{algorithmic}
\label{alg:detseq}
\end{small}
\end{algorithm}

Algorithm \ref{alg:detseq} simply loops over the vertices to find a set of weight-increasing $4$-cycles (denoted as $S$ in Algorithm~\ref{alg:detseq}). From these cycles $S$, a vertex disjoint set of weight-increasing cycles $D(S)$ is selected and then used to augment the matching in Line $10$. To find an alternating $4$-cycle that contains a vertex $j$ (also referred to as \textit{rooted at} $j$), select a neighbor $i\neq m_j$, then follow the matching edges incident to $i$ and $j$ to obtain $m_i$ and $m_j$. Since $\M$ is a perfect matching, these edges always exist. Next, scan the neighborhood of $m_j$ for $m_i$. If the edge $\{m_j,m_i\}$ exists, we have found an alternating $4$-cycle of gain $g(i,j,m_j,m_i)=w(\{i,j\})+w(\{m_i,m_j\})-w(\{i,m_i\})-w(\{m_j,j\})$. Clearly, this can be done in time $O(deg(m_j))$. To find the cycle with maximum gain rooted at $j$, we would have to repeat the process
$deg(j)-1$ times, resulting in a total time $O(deg(m_j)deg(j))$. However, if we instead mark the neighborhood of $m_j$ once, we can check for each neighbour $i$ of $j$ whether its mate $m_i$ was marked in constant time. Thus, for each vertex $j$, we can find the maximum gain $4$-cycle in $O(deg(m_i) + deg(j))$ time. When iterating over all vertices in this manner, every edge is visited twice, resulting in a linear $O(m)$ running time.

As mentioned above, Algorithm~\ref{alg:detseq} differs from the Pettie-Sanders algorithm~\cite{pettieSanders} in that it only finds cycles in Line $5$. Furthermore, the algorithms construct the vertex-disjoint set $D(S)$ in a different way. The Pettie-Sanders algorithm uses greedy selection to construct $S$. It sorts the elements of $S$ by gain in descending order, and then iterates over these augmentations in that order. Each augmentation from $S$ is added to the initially empty set $D(S)$, as long as it does not share a vertex with any augmentation already added to $D(S)$. For a sequential algorithm, this can be done in $O(n)$ time using topological sorting (see \cite{pettieSanders} for details). However, doing so in parallel would introduce a variable number of communication rounds and thus a significant load imbalance, and is therefore not suited for our purpose. Thus, we perform only local comparisons to find a heavy set of disjoint weight-increasing $4$-cycles. For the sequential case, this can be defined as follows: for each element $(i,j,m_j,m_i) \in S$, note its matched edges $\{j,m_j\}$ 
and $\{i,m_i\}$. Now, for each matched edge $e$ in arbitrary order, remove all except for the maximum gain cycle that contain it from $S$. Return the resulting $S$ as the disjoint set $D(S)$. In the worst case, this can result in applying only one augmentation per iteration. However, from a practical point of view such instances are extremely unlikely. See Section \ref{sec:algo1} for the parallel version of this strategy. 

The weight of the matching is then increased by augmenting $M$ with the elements of $D(S)$. The entire procedure is repeated until no more weight-increasing cycles can be found. In order to limit the running time in case many cycles of insignificant gain are found, we limit the repetitions by a small constant \textit{maxiters}. In our experiments, we set \textit{maxiters} to 10. However, for all problems that we experimented with, our algorithm terminates in less than 8 iterations.

}

\begin{figure}[!t]
    \centering
    \includegraphics[width=\textwidth]{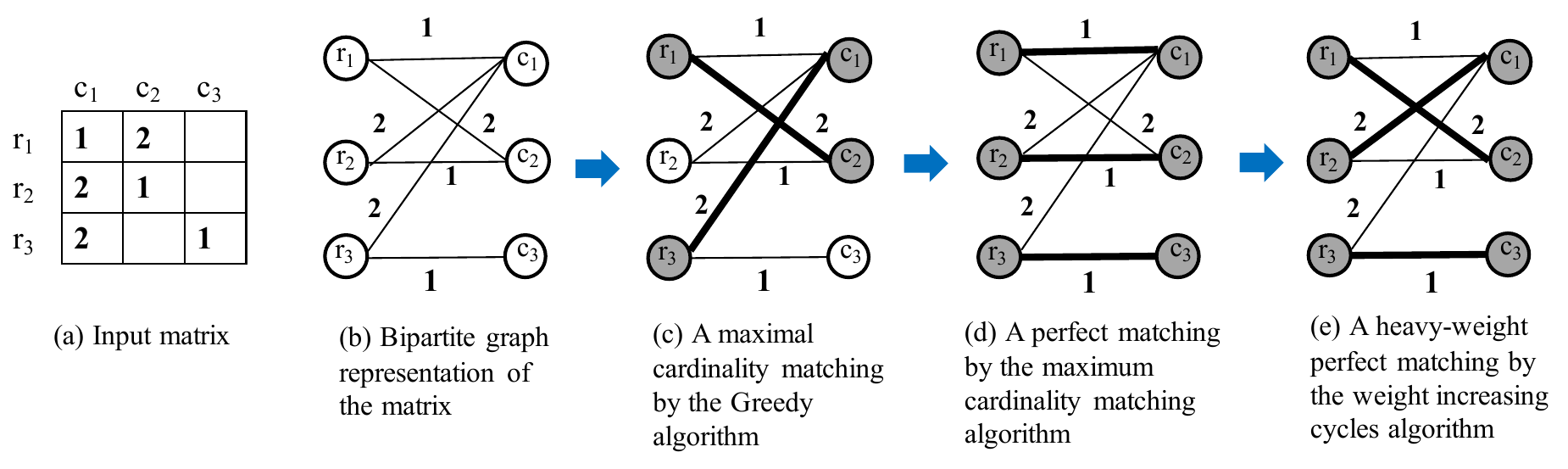}
    \caption{\lastrevision{The sequence of algorithms used to find a heavy-weight perfect matching in a bipartite graph. Matched vertices and edges are shown in dark shades. (a) A $3\times 3$ input matrix, (b) the bipartite graph representation of the matrix, (c) a maximal cardinality matching found by the Greedy algorithm, (d) a perfect matching obtained by the maximum cardinality matching algorithm (without using the tie-breaking heuristic discussed in Figure~\ref{fig:heuristic}), and (e) a heavy-weight perfect matching obtained by the weight increasing cycles algorithm. The algorithm in Subfig.~(d) is initialized by the matching in Subfig.~(c), and the algorithm in Subfig.~(e) is initialized by the matching in Subfig.~(d). In Subfig.~(d), the algorithm finds a weight-increasing cycle $(r_1, c_1, r_2, c_2, r_1)$, which is used to increase the matching weight in Subfig.~(e).}}
    \label{fig:HWPM-phases}
\end{figure}

\section{The Parallel Algorithm} 
\label{sec:algo1}
\subsection{Overview of the parallel HWPM algorithm}
\label{sec:cardalgo1}
\lastrevision{
The {\em heavy weight perfect matching}, or \texttt{HWPM} algorithm is a sequence of three matching algorithms.
The algorithm starts with a perfect matching obtained from a {\em maximum cardinality matching} (MCM) algorithm and improves the weight of the MCM by discovering weight-increasing cycles following the idea of the Pettie-Sanders algorithm discussed above.
The MCM algorithm is initialized using a maximal cardinality matching algorithm that returns a matching with cardinality at least half of the maximum. 
While this step is optional, doing so greatly decreases the running time of finding the MCM and also improves the parallel performance~\cite{duff2011design, azad2012multithreaded, AzadB16}.
Figure~\ref{fig:HWPM-phases} shows an example of the sequence of algorithms used to find a heavy-weight perfect matching in a bipartite graph.

Distributed memory algorithms for maximal and maximum cardinality matchings on bipartite graphs were developed in our prior work~\cite{matchingipdps16, AzadB16}. 
Among several variants of the maximal matching algorithms, we used a simple greedy algorithm to initialize MCM.
Our cardinality matching algorithms rely on a handful of bulk-synchronous matrix and vector operations.
For example, the MCM algorithm searches for \emph{augmenting paths} that alternate between matched and unmatched vertices with both endpoints being unmatched. 
This search for augmenting paths can be mapped to the sparse matrix-sparse vector multiplication (SpMSpV) for which efficient distributed-memory parallel algorithms were developed~\cite{matchingipdps16}. 
For this work, we used a new heuristic with the MCM algorithm so that the initial perfect matching has higher weights.
As we will show in the result section, the heuristic improves the weight of the final matching for many input graphs. 





\begin{figure}[!t]
    \centering
    \includegraphics[scale=.4]{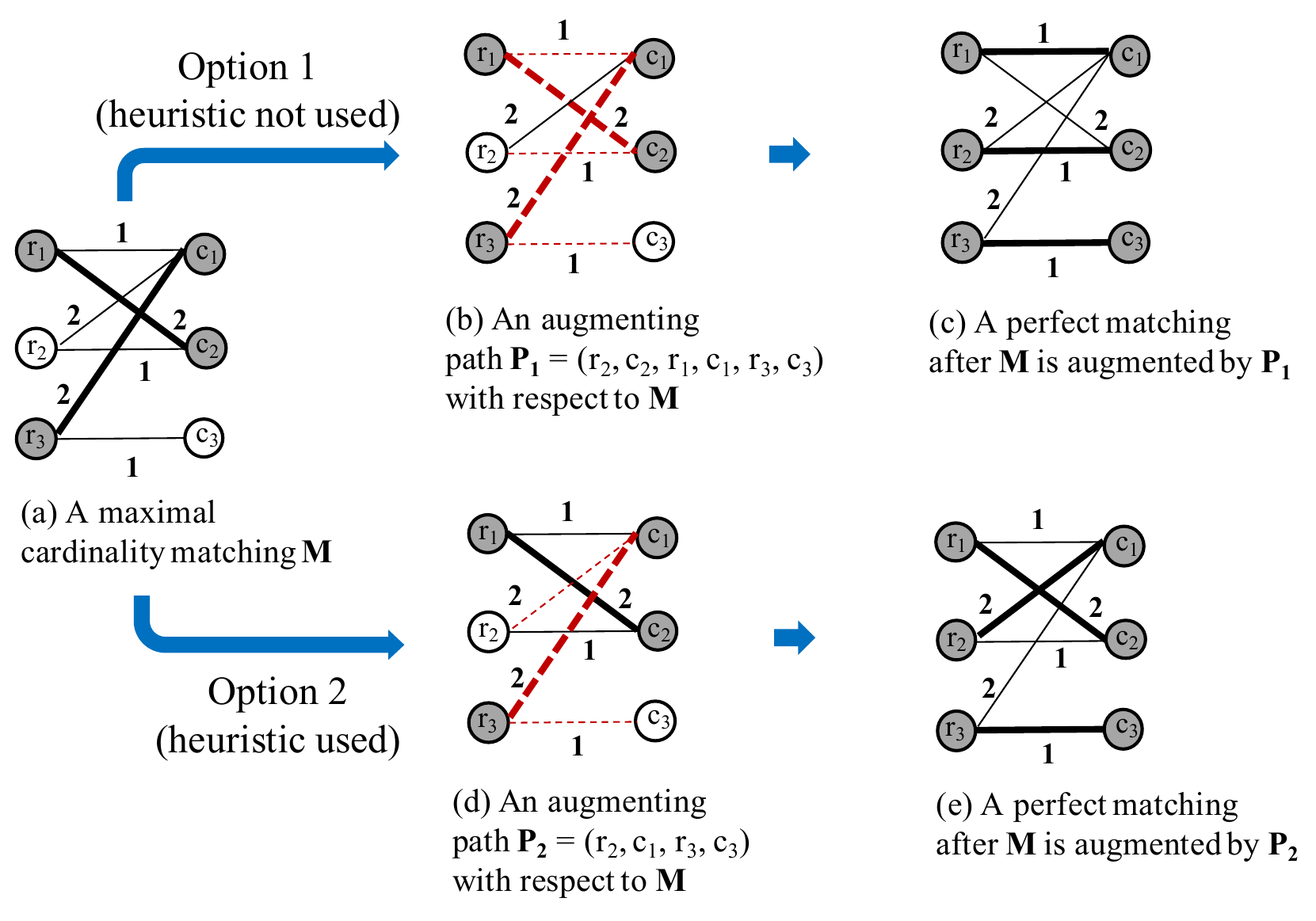}
    \caption{\lastrevision{An example of using the tie-breaking heuristic in the MCM algorithm. Matched vertices and edges are shown in dark shades, and augmenting paths are shown in dashed lines. Given a maximal matching $M$ in Subfigure~(a), we can discover two augmenting paths $P_1$ shown in Subfigure~(b) and $P_2$ shown in Subfigure~(d). Between these two augmenting paths, $P_2$ uses the tie-breaking heuristic because it has high local weights starting from the vertex $r_2$. Subfigure~(c) and Subfigure~(e) show two perfect matchings obtained by augmenting $M$ by $P_1$ and $P_2$, respectively. Subfigure~(c) contains a weight-increasing cycle $(r_1, c_1, r_2, c_2, r_1)$ that can be used to improve the perfect matching. By contrast, the perfect matching in Subfigure~(e) does not contain any weight-increasing cycle.}}
    \label{fig:heuristic}
\end{figure}

\subsection{Tie-breaking heuristic for the MCM algorithm}
The MCM algorithm searches for augmenting paths from unmatched vertices and uses these paths to increase the cardinality of the matching.
If there are multiple augmenting paths from the same source vertex, any of the paths can be used for maximum cardinality.
For this work, we modified the original cardinality matching algorithms in such a way that when selecting potential matching edges, we break ties by giving precedence to edges with higher weight.
Figure~\ref{fig:heuristic}(b) and (d) show that there are two augmenting paths starting from vertex $r_2$ in Subfigure~\ref{fig:heuristic}(a).
If the algorithm uses the heavy-weight-tie-breaking heuristic, it finds the path in  Subfigure~\ref{fig:heuristic}(d) and the perfect matching shown in Subfigure~\ref{fig:heuristic}(e).
Since the weight of the matching in Subfigure~\ref{fig:heuristic}(e) is larger than the matching in Subfigure~\ref{fig:heuristic}(c), the former can improve the quality of the final heavy-weight perfect matching and reduce the work needed in the last weight-increasing-cycles algorithm.
For example, the matching in Subfigure~\ref{fig:heuristic}(e) does not contain any weight-increasing cycles; hence, the last matching step finishes after just one iteration, which can greatly reduce the total running time of our algorithm.

This simple heuristic often results in the perfect matchings having high weight without incurring any additional cost. 
As mentioned in the previous paragraph, our algorithm is based on linear algebra operations such as SpMSpV. In these operations, we can replace traditional ``addition" and ``multiplication" operators with other user supplied semiring operators~\cite{kepner2011graph}. In graph traversal terms, it is helpful to think of this semiring ``addition" operator as a means of choosing among multiple possible paths. 
For our heuristic, we use ``max" as our semiring ``addition" operator so that edges with higher weights are selected when multiple options are available in the MCM algorithm.
Replacing individual arithmetic operations does not change the computational pattern of the MCM algorithms.
Hence, the heuristic improves the quality of the final matching without adding additional computational costs. 

The tie-breaking heuristic is also applied in the maximal matching algorithm in the first step. 
In that algorithm, we select edges with heavy weights when greedily selecting edges for matching.
Similar to the MCM algorithm, the tie-breaking heuristic does not incur any additional cost for the maximal matching.
Finally, if we obtain a perfect matching from the maximal matching algorithm, we do not run the MCM algorithm at all.
All these heuristics and optimizations help us obtain a good perfect matching quickly.
Once a perfect matching is obtained, we improve its weight using our newly developed parallel algorithm. 
}

\subsection{The weight increasing alternating cycles algorithm}
As mentioned in the last section, the algorithm for maximum weight approximation aims to find weight-increasing $4$-cycles. Unlike longer cycles, they can be found in time proportional to the degree of their vertices, and doing so requires only a constant number of communication rounds in the distributed memory parallel setting. Thus, our algorithm can be considered a special case of the general \emph{augmenting cycles algorithm}\footnote{The name is established in the literature. However, we reserve the word \emph{augmenting} for cardinality-increasing paths and otherwise use the term \emph{weight-increasing} cycles and paths.}. We will refer to it as the \texttt{WIAC} for \textit{weight increasing alternating cycles} algorithm.

\begin{figure}[t]
    \centering
    \includegraphics[width=0.35\textwidth]{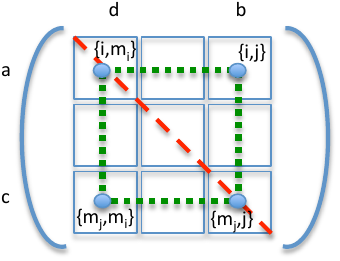}
    \caption{$4$-cycle in the 2D distributed matrix. Each square contains the part of the matrix assigned to one process. For simplicity, the matched edges are gathered on the main diagonal (dashed red line). $a$ and $c$ denote rows of processes, while $b$ and $d$ denote columns. Thus, if edge $\{i,j\}$ is on process $(a,b)$, then its incident matched edges are on process $(a,d)$ and $(c,b)$ respectively, where $c$ and $d$ depend on the position of the matched edges.  If the unmatched edge $\{m_j,m_i\}$ exists in the graph, it must be on process 
    $(c,d)$.}
    \label{fig:matrixcycle}
\end{figure}

The vertices of a $4$-cycle might be distributed over 4 processes, requiring 4 major communication steps, which will be described in detail below. We use a regular 2D partitioning of the input matrix. Thus, $p$ processes form a $\sqrt{p}\times\sqrt{p}$ grid. Processes are indexed by their position in the grid and denoted by \textit{process} $(row,column)$. 
Figure \ref{fig:matrixcycle} illustrates the layout, and explains the labeling of the processes. 

The cardinality algorithm stores the matching it computed in two vectors (one containing the matching partners for the columns, one for the rows) which are distributed among all processors. 
However, finding weight-increasing cycles efficiently requires replicating parts of these vectors
on each processor, along with the weight of the matched edges. Thus, the algorithm starts by distributing the matching vectors across rows and columns using broadcasts, to be stored in local arrays for fast access. After initialization, the algorithm loops over the four fundamental steps $maxiters$ times, or until no more weight-increasing $4$-cycles can be found, as shown in Algorithm \ref{alg:PPRoverview}.

\begin{algorithm}[t]
\begin{small}
\caption{Sketch of the Basic Parallel Algorithm }
\begin{algorithmic}[1]
\State \textbf{Input}: a weighted graph $G=(R \cup C, E, w)$ 
\State \textbf{On each process $(a,b)$ do in parallel:}
	\ForAll{rows $i$ of $\Am$ assigned to process $(a,b)$} 
		\State Receive $m_i$ and $w(i,m_i)$ and store them in a local array 
	\EndFor
	\ForAll{columns $j$ of $\Am$ assigned to process $(a,b)$}
		\State Receive $m_j$ and $w(m_j,j)$ and store them in a local array 
	\EndFor	
\For{i = 1 to $maxiters$}
	\State Step A: Generate requests for cycles
	\State Step B: Check whether cycles of positive gain exist
	\State Step C: Determine winning cycle for each edge $\{m_j,j\}$
	\State Step D: Determine winning cycle for each edge $\{i,m_i\}$ and change the matching
    \If{no cycle was found} 
    	\State \textbf{break}
    \EndIf
\EndFor
\end{algorithmic}
\label{alg:PPRoverview}
\end{small}
\end{algorithm}

Like the deterministic sequential Algorithm \ref{alg:detseq}, we construct a set of vertex-disjoint cycles. In effect, we parallelize the \textbf{FOR ALL} statement in Line 4 of Algorithm \ref{alg:detseq}. However, as mentioned in Section~\ref{sec:algoPS}, we do not use the same greedy strategy as the sequential algorithm to pick the weight-increasing $4$-cycles in Line 7, since the standard greedy algorithm is inherently sequential. Instead, we perform limited local weight comparisons, which are described below. 

After initialization, the first step will be to select vertices at which the cycles are rooted. While all vertices are eligible, we can reduce this number since each cycle can be rooted at any of its 4 vertices. Because the matrix is stored in Compressed Sparse Column (CSC) format, we only start cycles from column vertices, reducing the number of root vertices by half. 
\lastrevision{
This number can be further reduced by selecting next row vertices whose indices are higher than the mate of the initial column vertex.}


Now, for a potential cycle rooted at a column vertex $j$ and containing a row vertex $i$, we generate a request to the owner of $\{m_j,m_i\}$, which includes the edge weight
$w(\{i,j\})$, as shown in Algorithm \ref{alg:PPRoverviewA}. We label these 3-tuples \textit{A-requests}. For performance reasons, the exchange of all such requests is bundeled into an All-to-All collective operation.

\begin{algorithm}[t]
\begin{small}
\caption{Step A - Generate requests for cycles}
\begin{algorithmic}[1]
\ForAll{processes ($a,b$)} 
	\ForAll{columns $j$ of $\Am$ assigned to process $(a,b)$} 
		\ForAll{rows $i>m_j$ of $\Am$ assigned to process $(a,b)$}
         	\If{\{$i$,$j$\} exists} 
				\State Let $(c,d)$ be the owner of $\{m_j,m_i\}$ 
                \State \parbox[t]{\dimexpr\linewidth-6em}{Add A-request$(m_j,m_i,w(\{i,j\})$ to request queue for process $(c,d)$\strut} 
    		\EndIf
		\EndFor
	\EndFor
\EndFor	
\State Exchange A-requests via AllToAll communication
\end{algorithmic}
\label{alg:PPRoverviewA}
\end{small}
\end{algorithm}

In the next step, we need to determine if $\{m_j,m_i\}$ and thus the alternating cycle of $i,j,m_j,m_i$ exists and is weight-increasing (i.e.,~it has $g(i,j,m_j,m_i)>0$). Note that the weight of each matching edges is stored on all processes in the same grid row/column, and can thus be accessed directly, as shown in Algorithm \ref{alg:PPRoverviewB}. 
However, before we can flip the matching along this cycle, we have to make sure that it is vertex disjoint with other candidate cycles, and if not, find out which cycle has the highest gain. We therefore send a request to the process that owns the matched edge $\{m_j, j\}$.

\begin{algorithm}[t]
\begin{small}
\caption{Step B - Check whether cycles of positive gain exist}
\begin{algorithmic}[1]
\ForAll{processes $(c,d)$:} 
	\ForAll{A-requests$(m_j,m_i,w(\{i,j\})$ from $(a,b)$}
		\If{$\{m_j,m_i\}$ exists:}
            \State \parbox[t]{\dimexpr\linewidth-6em}{$g(i,j,m_j,m_i)=w(\{i,j\})+w(\{m_i,m_j\})-w(\{i,m_i\})-w(\{m_j,j\})$\strut}   
			\If{$g(i,j,m_j,m_i) > 0$}
            	 \State \parbox[t]{\dimexpr\linewidth-6em}{Add B-request $(i,j,m_j,m_i,g(i,j,m_j,m_i))$ to request queue for process $(c,b)$\strut} 
			\EndIf
		\EndIf	
	\EndFor
\EndFor	
\State Exchange B-requests via AllToAll communication 
\end{algorithmic}
\label{alg:PPRoverviewB}
\end{small}
\end{algorithm}

\begin{algorithm}[!t]
\begin{small}
\caption{Step C - Determine winning cycle for each edge $\{m_j,j\}$}
\begin{algorithmic}[1]

\ForAll{processes $(c,b)$:}
	\ForAll{rows $m_j$ of $\Am$ assigned to process $(c,b)$} 
        \State \parbox[t]{\dimexpr\linewidth-4.5em}{Find B-request $(i,j,m_j,m_i,g(i,j,m_j,m_i))$ with maximum gain\strut}
        \State \parbox[t]{\dimexpr\linewidth-4.5em}{Add C-request $(i,j,m_j,m_i,g(i,j,m_j,m_i))$ to request queue for process $(a,d)$ \strut}
	\EndFor
\EndFor				
\State Exchange C-requests via AllToAll communication

\end{algorithmic}
\label{alg:PPRoverviewC}
\end{small}
\end{algorithm}

\begin{figure}
    \centering
    \includegraphics[width=0.48\textwidth]{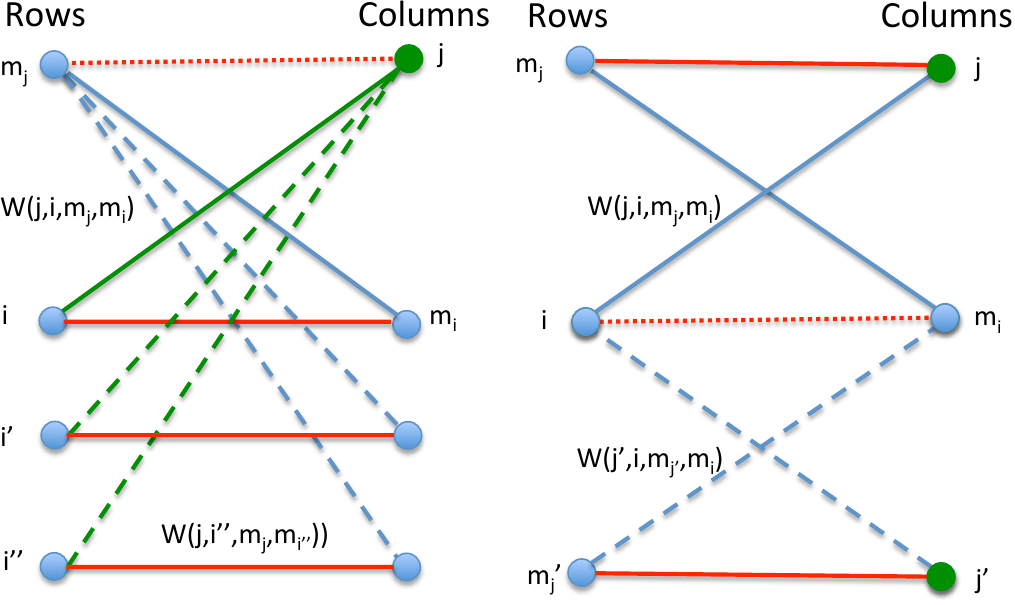}
    \caption{Cycle collisions in the graph view. Left: multiple cycles, all of which are rooted in column vertex $j$ compete for the matched edge $\{m_j,j\}$ in Step C of the algorithm. Right: cycles rooted at different column vertices $j$ and $j'$ compete for the matched edge $\{i,m_i\}$ in Step D.}
    \label{fig:phase34}
\end{figure}

Now, the owner of each edge $\{m_j,j\}$ collects all incoming requests for that edge, selects one with maximum gain, and discards all others. Since these requests correspond to cycles rooted at $j$, all remaining cycles are now disjoint w.r.t.~their $\{m_j,j\}$ edge, as shown in Algorithm \ref{alg:PPRoverviewC}. However, we still have to 
ensure that the cycles are disjoint with other cycles at the edge $\{i,m_i\}$. Therefore, we send a request to its owner, which might not share any column or row with the sending process.  Figure \ref{fig:phase34} shows an example of such competing cycles.

\begin{algorithm}[!t]

\begin{algorithmic}[1]
\begin{small}
\ForAll{processes $(a,d)$:}
	\ForAll{columns $j$ of $\Am$ assigned to process $(a,d)$}
		\If{No C-request was sent from $i$ in Step C }	
          \State \parbox[t]{\dimexpr\linewidth-4.5em}{Find C-request$(i,j,m_j,m_i,g(i,j,m_j,m_i))$ with  maximum gain \strut}
		  \State $k$ = $m_i$
		  \State $l$ = $m_j$
		  \State Broadcast $m_i=j, w(i,j)$ to all processes $(a,*)$
		  \State Broadcast $m_j=i, w(i,j)$ to all processes $(*,b)$
		  \State Broadcast $m_k=l, w(l,k)$ to all processes $(c,*)$
		  \State Broadcast $m_l=k, w(l,k)$ to all processes $(*,d)$
		\EndIf				
	\EndFor
\EndFor	
\end{small}
\end{algorithmic}

\caption{Step D - Determine winning cycle for each edge $\{i,m_i\}$ and change the matching}

\label{alg:PPRoverviewD}

\end{algorithm}
 
\begin{figure}[t]
    \centering
    \includegraphics[width=0.35\textwidth]{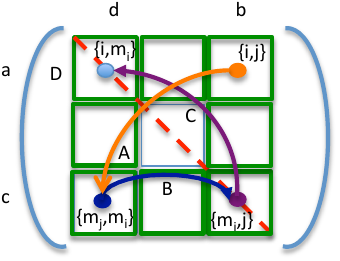}
    \caption{Communication in the 2D distributed matrix view. In Step A (orange arrow) communication goes from $(a,b)$ to $(c,d)$. Step B (blue arrow) from $(c,d)$ to $(a,d)$, and Step C (purple arrow) from $(c,b)$ to $(a,d)$. In Step D, the matching is updated along rows $a$ and $c$, and along columns $b$ and $d$ (green squares). In this example that includes all but the process in the center.}
    \label{fig:matrixcycle2}
\end{figure}

In Step D, the owner of each edge $\{i,m_i\}$ collects requests for that edge, selects one with maximum gain, and discards all others, similar to Step C. Thus, all cycles are disjoint w.r.t.~their $\{i,m_i\}$ edges. However,
it is still possible that the $\{i,m_i\}$ edge of one cycle is the $\{m_j,j\}$ edge of a different cycle. Thus, if a process sent a C-request for an edge $e=\{m_j,j\}$ in Step C, then it will automatically discard the requests for other cycles that have $e$ as their $\{i,m_i\}$ edge. As mentioned in Section \ref{sec:algoPS}, our strategy deviates from the Pettie-Sanders algorithm here. The reason for this lies in the fact that finding the maximal set of weight-increasing $4$-cycles would incur additional communication that most likely affects only a small number of vertices. Therefore, in the parallel case it is preferable to simply drop the problematic cycles and generate new ones instead. 

The final step consists of flipping matched and unmatched edges in each cycle and communicating the change along the rows and columns, which is shown in Algorithm \ref{alg:PPRoverviewD}. The broadcast operations simply inform all processes in the two affected rows and columns that the matched edges and thus vertices have changed, and causes all affected processes to update their matching information, along with the associated weight.
The entire path of the communication is sketched in Figure  \ref{fig:matrixcycle2}.


\subsection{Analysis of the parallel HWPM algorithm}

We measure communication by the number of {\em words} moved ($\W$) and
the number of {\em messages} sent ($S$). The cost of communicating a $\W$-word message is $\alpha + \beta \W$ where $\alpha$ is the
latency and $\beta$ is the inverse bandwidth, both are defined relative to the cost of a single arithmetic operation. Hence, an algorithm that
performs $F$ arithmetic operations, sends $S$ messages, and moves $\W$ words takes $F + \alpha S + \beta \W$ time. 

\finalrevision{In our experiments, we balance load across processes by randomly permuting rows and columns of the input matrix  before running the matching algorithm. We distribute our matrices and vectors on a 2D $\sqrt{p} \times \sqrt{p}$ process grid. First applying a random row and column permutation, and then block distributing the matrix on a 2D process grid is known to provide surprisingly good load balance for large sparse matrices in practice~\cite{borvstnik2014sparse, bulucc2012parallel}. For the restricted case where no row or column is ``dense'', Ogielski and Aiello~\cite{ogielski1993sparse} proved that the probability of having a processor with more than $(1+\epsilon)^2 \mathit{nnz} / p$ nonzeros is less than $e^{-O(\epsilon h(\epsilon))}$ where $h(x)$ is a strictly increasing function with $h(x)\approx x/2$ for $x \to 0$ and $h(x)\approx \ln(x)$ for $x \to \infty$.}

\finalrevision{In practice, even a symmetric permutation, where rows and columns are applied the same permutation, provides decent load balance as we demonstrate experimentally in Section~\ref{sec:load-balance}. Based on Ogielski and Aiello's theoretical justification for the restricted case and the experimental results supporting the theory on matrices satisfying fewer assumptions, we initially assume in our analysis that matrix nonzeros are independently and identically distributed. Later in this section, we will drop the restriction of matrices having no dense rows or columns.}

The running time of the MCM algorithm is dominated by parallel sparse matrix-sparse vector multiplication whose complexity has been analyzed under the \iid~assumption in our prior work~\cite{matchingipdps16}.
Let $\left\vert{\mathrm{iters_{MCM}}} \right\vert$ to be the number of iterations needed to find an MCM, the expected complexity of the MCM step is:
$$ T_{\text{MCM}}  = O \Bigl( \frac{m}{p} +  \beta  \bigl ( \frac{m}{p} + \frac{n}{\sqrt{p}} \bigr ) +  \left\vert{\mathrm{iters_{MCM}}} \right\vert \alpha \sqrt{p} \Bigr) .$$



 In Step A of the \texttt{WIAC} algorithm, we exchange a total of $O(m)$ requests among all processes. Under the \iid~assumption, each process contains $m/p$ edges and spends $O(m/p)$ time to prepare $O(m/p)$ requests for all other processes. 
 Let $R_i$ and $C_i$ be the set of rows and column whose pairwise connections are stored in process $i$. 
 Let $\M(R_i)$  and $\M(C_i)$ be the set of mates of vertices in $R_i$ and $C_i$, respectively.
 Since a process stores at most $n/\sqrt{p}$ rows and columns, $|\M(R_i)| \leq n/\sqrt{p}$ and $|\M(C_i)| \leq n/\sqrt{p}$.
 In order to receive a request in the $i$th process, there must be an edge between a pair of vertices in  $\M(R_i) \times \M(C_i)$. 
 Under the \iid~assumption, the number of such edges is $O(m/p)$.
 Hence a process receives $O(m/p)$ messages in the communication round of Step A. 

 The number of requests in Step B cannot be greater than in Step A,
 and requests in Step C and matching operations in Step D are bounded by $O(n)$. 
 Hence the total cost of \texttt{WIAC} under the \iid~assumption is: 
 $$ T_{\text{WIAC}}  = O \Bigl( \left\vert{\mathrm{iters_{WIAC}}} \right\vert\Bigl(\frac{m}{p} +  \beta \frac{m}{p} +  \alpha p \Bigr)\Bigr) .$$
 Note that in \texttt{WIAC}, $\left\vert{\mathrm{iters_{WIAC}}} \right\vert$ is bounded by a small constant $\mathit{maxiters}$ which serves to stop the algorithm when too many successive cycles of small gain are found. Therefore, we will drop it in the general (not \iid ) analysis below.

\finalrevision{Now we will relax the assumption of no rows or columns being dense and update our analysis accordingly. For a graph with $n$ vertices represented as a square $n\times n$ sparse matrix, let us define the ratio of edges to vertices by $d = \mathit{nnz}/n$. By definition, there can be at most $O(d)$ dense rows and columns where a dense row is a row that has $O(n)$ nonzeros. We can apply the balls into bins analysis by considering each dense row as a ball and each process row as a bin. When the number of bins is equal to the number of balls $b$, a well-known bound on the maximum number of balls assigned to a single bin is $\ln{b}/\ln{\ln{b}}$ with high ($\approx 1-1/b$) probability~\cite{mitzenmacher2017probability}. The same bound naturally holds as we increase the number of bins while keeping the number of balls $b$ fixed.}

\finalrevision{
\begin{theorem}
Given an $n\times n$ sparse matrix with $\mathit{nnz}$ nonzeros, we first apply random row and column permutations. We then
distribute the matrix in a 2D block fashion on a $\sqrt{p} \times \sqrt{p}$ process grid. 
The maximum number of nonzeros per processor is 
$$ \frac{\ln{d}}{\ln{\ln{d}}} \frac{n}{\sqrt{p}} + (1+\epsilon)^2 \frac{\mathit{nnz}}{p}
$$
with high probability when $\sqrt{p} \geq d = \mathit{nnz}/n$.
\end{theorem}}

\finalrevision{
\begin{proof}
For $\sqrt{p} \geq d$, the maximum number of dense rows assigned to a process row is $\ln{d}/\ln{\ln{d}}$ with high probability. Since a process row has $\sqrt{p}$ processes, each member gets at most $\ln{d}/\ln{\ln{d}} \cdot n / \sqrt{p}$ nonzeros due to its share from dense rows. Combining this with the results of Ogielski and Aiello that show the part of the matrix without dense rows and columns is well-balanced with $(1+\epsilon)^2 \mathit{nnz}/p$ nonzeros per process with high probability, we achieve the desired bound. 
\end{proof}}

\finalrevision{
Hence, in the case where the matrix contains dense rows, the expected (with high probability) running time of our full algorithm becomes:
\begin{equation} 
 T_{\text{HWPM}} = O \Bigl((1+ \beta) \bigl( \frac{\ln{d}}{\ln{\ln{d}}} \frac{n}{\sqrt{p}} + (1+\epsilon)^2 \frac{m}{p} \bigr) +  \alpha \bigl(p +  \left\vert{\mathrm{iters_{MCM}}} \right\vert \sqrt{p} \bigl)\Bigr).
 \end{equation}}
 
 \finalrevision{
We presented this bound in terms of the number of edges $m$ of the graph instead of the number of nonzeros $\mathit{nnz}$ of its associated matrix because the bound concerns the running time of a parallel graph algorithm. Note that the iteration count is independent of the number of processes used. The good news is that in the regime where the performance is bounded by the inter-process bandwidth, the performance of our algorithm is expected to scale with $\sqrt{p}$ in the worst case with high probability and with $p$ in the best case. The bad news is that once the performance starts being bounded by latency, we should not expect any speedups as we add more processes. In the case of extremely large $p$, we might even experience slowdowns in strong scaling.}

{
\setlength{\tabcolsep}{5pt}
\begin{table}[!t]{
\centering

\caption{Overview of Evaluated Platforms.  $^1$Shared between 2 cores in a tile. 
}

\begin{tabular}{rcc}
					&  {\bf Cori} & {\bf Edison }    \\
					
\hline
{\bf Core }	 	 & Intel KNL			&  Intel Ivy Bridge\\

\hline
Clock (GHz)			& 1.4			& 2.4					\\
L1 Cache (KB)		& 32		& 32				\\
L2 Cache (KB)		& 1024$^1$		& 256				\\
\hline
{\bf Node Arch.}	 	 & 			& \\
\hline
Sockets/node			&  1		&	2					\\
Cores/socket			& 68				& 12					\\
Memory (GB)		&  96	&	64			\\
\hline
{\bf Overall system}	 	 & 			& \\
\hline
Nodes & 9,688 & 5,586\\
Interconnect & Aries (Dragonfly)	&  Aries (Dragonfly)\\

\hline
{\bf Prog. Environment}	 	 & 			& \\
\hline
Compiler & \multicolumn{2}{c}{Intel C++ Compiler (icpc) ver18.0.0}  \\
Optimization & -O3 &  -O3 \\
\hline
\end{tabular}

\label{tab:machines}
}
\end{table}
}

\section{Results}
\label{sec:results}

\subsection{Experimental setup} 
\label{sec:setup}
We evaluated the performance of our algorithms on the Edison and Cori supercomputers at NERSC.
On Cori, we used KNL nodes set to cache mode where the available 16GB MCDRAM is used as cache. 
Table~\ref{tab:machines} summarizes key features of these systems. 
We used Cray's MPI implementation for inter-node communication and OpenMP for intra-node multithreading. The multithreading is used to parallelize all \textbf{for all} statements in Algorithm \ref{alg:PPRoverview} through \ref{alg:PPRoverviewD}.
Unless otherwise stated, all of our experiments used 4 MPI processes per node, and thus 6 (Edison) or 16 (Cori) OpenMP threads per process. This configuration was found to be superior for the cardinality matching algorithm~\cite{matchingipdps16} which tends to dominate the running time for large core counts.
We rely on the Combinatorial BLAS (CombBLAS) library~\cite{CombBLAS} for parallel file I/O, data distribution and storage. 
We always used square process grids because rectangular grids are not supported in CombBLAS. 


\lastrevision{
We experimented with over 100 sparse matrices from the SuiteSparse matrix collection~\cite{ufget} and from other sources. From SuiteSparse, we selected unsymmetric matrices that have more than 5,000 rows and columns. For small matrices, we only show the quality of matchings from MCM, HWPM and MC64 algorithms. We show their matching weights, obtained approximation ratios, and their performance when used with SuperLU\_DIST. 
We selected a set of large matrices with at least 10 million nonzeros 
that have a different average number of nonzeros per column and diverse nonzero patterns. The properties of these large matrices are shown in Table~\ref{table:problem-statistics}.
The details of the smaller matrices used in Table~\ref{table:weights-all} can be found in the SuiteSparse matrix collection (https://sparse.tamu.edu/).
Explicit zero entries are removed from the matrices in the preprocessing step. 
Before running matching algorithms, we equilibrate the matrices by diagonal scaling so that the  maximum entry in magnitude of each row or column is $1$. This usually reduces the condition number of the system.

}

We compare the performance of our parallel algorithm with the MWPM  implementations in MC64~\cite{MC64} and an optimized sequential implementation of HWPM.
Our sequential implementation uses the Karp-Sipser and Push-Relabel algorithms~\cite{KLMU2012} to find a perfect matching, and then uses the weight-increasing alternating cycles algorithm to find an HWPM. 
\finalrevision{MC64 provides five options for computing a matching, while HWPM provides two.
At the same time, there are three relevant objectives. The first is to maximize the sum of weights on matched edges. MC64 provides this as Option 4, and HWPM as Option A. We use this option for running time and approximation quality comparisons. The second objective is to maximize the product of weights on matched edges. This is HWPM Option B. We use this objective to study the impact of the matchings on linear solver accuracy. Since the product can be maximized by maximizing the sum of logarithms, there is little algorithmic difference between this and the first objective. MC64 does not directly provide this, but it can be replicated by giving the logarithm of the weights as input.
As a third alternative, MC64 offers Option 5 which maximizes their product and then uses the dual variables for scaling. Since HWPM does not produce dual variables, we do not use this option for comparison.}
Since a matching algorithm is often used as  a subroutine of another application, we exclude file I/O and graph generation time when reporting the running time of matching algorithms.

\begin{table}[!t]
 \centering
 \caption{Square matrices used to evaluate matching algorithms. All matrices, except \textit{Li7Nmax6, perf008cr, Nm7, NaluR3,} and \textit{A05} are from the SuiteSparse matrix collection~\cite{ufget}. Explicit zero entries are excluded beforehand. 
We separate larger matrices ($nnz >$ 200M) by a horizontal line.
}
 \begin{tabular}{@{} l l r  r  r r l   @{}}
    \toprule
    Matrix &	columns &	nnz & 	avg nnz & max nnz \\
 & 	($\times 10^6$)	&	($\times 10^6$)	& per column & per column &	Description\\
 \toprule

memchip	&	2.71	&	13.34 & 4.92 & 	27 &	circuit simulation  \\
Freescale2	&	3.00	&	14.31	& 4.77 & 14,891 &	circuit simulation \\
rajat31	&	4.69	&	20.32	& 4.33 & 1,252 &	circuit simulation  \\

boneS10	&	0.91	&	40.88	& 44.92 & 67 &	model reduction problem \\
Serena		&		1.39	&	46.14	& 23.71 & 249 &	gas reservoir simulation \\
circuit5M	&		5.56	&	59.52 	& 10.70 & 1,290,501 & circuit simulation \\
audikw\_1	&	0.94	&	77.65	& 82.60 & 345 &	structural prob \\
dielFilterV3real	&	1.10	&	89.31	& 81.19 & 270 & higher-order finite element \\
Flan\_1565	&	1.56	&	114.17	& 73.19 & 81 &	structural problem \\
\midrule
HV15R	&		2.02	&	283.07	& 140.13 & 303 &	3D engine fan	\\
Li7Nmax6~\cite{aktulga2014optimizing}	&		0.66	&	421.94	& 639.30 & 5,760 &	nuclear config. interaction \\

Nm7~\cite{aktulga2014optimizing}	&		4.08	&	437.37	& 107.20 & 1627 & 	nuclear config. interaction \\
NaluR3~\cite{lin2014towards}	&		17.60	&	473.71	& 26.92 & 27 & 	Low Mach fluid flow\\
perf008cr	& 7.90	& 634.22 & 80.28 & 81 & 3D structural mechanics \\
nlpkkt240	&	27.99	&	760.65	& 27.18 & 28 &	Sym. indef. KKT matrix \\
A05	& 	1.56	& 1088.71 & 697.89 &  1,188 & MHD for plasma fusion \\

      \toprule  
      \end{tabular}
\label{table:problem-statistics}
 \end{table}
 



 
\subsection{Approximation ratio attained by parallel HWPM algorithm} 
\lastrevision{
Table~\ref{table:weights-all} shows the quality of matchings from MC64, MCM and HWPM algorithms for all small-scale matrices. 
Interpretation of the columns are explained in the caption of the table. 
We compute the approximation ratio by dividing the weight of MCM or HWPM by the optimum weight computed by MC64 and show it as a percentage.  
Even though our parallel algorithm can not guarantee the theoretical bound of the Pettie-Sanders algorithm, the obtained matchings are often very close to the optimum in practice. 
There is only one matrix (tmt\_unsym) where HWPM achieves an approximation ratio of 25\%. The minimum approximation ratio from HWPM (with heuristic) for all other matrices is 84.46\%.
This result demonstrates that HWPM is very successful in obtaining near-optimum perfect matchings for almost all practical problems. 
However, examples like tmt\_unsym still exist since we can not provide an approximation guarantee. 
In SuiteSparse matrix collection, there are matrices that are already in perfect matching with maximum weight state (matrices in the last part of Table~\ref{table:weights-all}). MCM may easily find MWPM for these matrices as shown in the last part of Table~\ref{table:weights-all}. 
We still keep these matrices in our result because one still has to run MC64 or HWPM to make sure that it is indeed a maximum weight matching.

{\bf Impact of the tie-breaking heuristic.} As mentioned in Section~\ref{sec:cardalgo1}, the HWPM algorithm starts with a perfect matching obtained from an MCM algorithm.  
The MCM algorithm may use a heuristic that picks heavy-weight edges when breaking ties.
Table~\ref{table:weights-all} shows that this heavy-weight-tie-breaking heuristic generally improves the weights of the final HWPM for about 70\% matrices in our test suite.
When the tie-breaking heuristic is not used, MCM's weights can be small because the MCM algorithm ignores edge weights without the heuristic. As a result, MCM without the heuristic provides lower approximation ratios (average 73.10\% and minimum 18.84\%). 
In this case, HWPM's weights are significantly better than MCM's weights because HWPM has more room for weight improvement. Column 5 in  Table~\ref{table:weights-all} shows that HWPM's approximation ratio can be up to 77.31\% higher than MCM's approximation ratio. Overall, the approximation ratios of HWPM without heuristic are on average 91.95\%, with the minimum lies at 25\%.

\footnotesize
\begin{center}
\begin{longtable}[H]{@{} |l r | r r r | r r   r  | r |@{}}
 \caption{\lastrevision{The quality of matchings from MC64, MCM and HWPM with and without the heavy-weight tie breaking heuristic. As mentioned in Section~\ref{sec:setup}, equilibration is used before we compute the matching.
 Column 2 reports the MWPM weights obtained from MC64. Columns 3 and 4 show the approximation ratios obtained by MCM and HWPM algorithms without the heavy-weight tie breaking heuristic. Columns 6 and 7 show the approximation ratios obtained by MCM and HWPM algorithms with the heuristic. Column 5 and 8 show the difference between their previous two columns (denoting the improvement from the weight increasing alternating cycles algorithm). Column 9 shows the overall improvement in HWPM when the heuristic is used, which is calculated by subtracting col 4 from column 7. We sort rows in descending order of the last column. Average, minimum and maximum approximation ratios are shown in the last three rows.
 The following matrices are not shown because MCM and HWPM both attain 100\% approximation ratio: 
 ML\_Laplace, ASIC\_100ks, ASIC\_680ks, FEM\_3D\_thermal2,ML\_Geer, bcircuit, cage12, cage13, circuit5M\_dc, dc1, dc2, ecl32, epb3, poisson3Db, stomach, torso2, torso3, trans4, trans5, venkat01, venkat50, water\_tank, xenon2, dc3.
 }}\\

\hline 
  
    &  \textbf{MC64} & \multicolumn{3}{c |}{\textbf{Without Heuristic}} & \multicolumn{3}{c |}{\textbf{With Heuristic}}  &	\textbf{HWPM} \\ [1.1ex]
    \textbf{Matrix} & \textbf{Weight}	&	$\frac{\textbf{MCM}}{\textbf{MC64}}$ & $\frac{\textbf{HWPM}}{\textbf{MC64}}$ &  
     \multicolumn{1}{c|}{\textbf{Diff}} & 
     $\frac{\textbf{MCM}}{\textbf{MC64}}$ &
     $\frac{\textbf{HWPM}}{\textbf{MC64}}$ &
     \multicolumn{1}{c|}{\textbf{Diff}} &
     \textbf{Improv.}\\[1.1ex]

\endfirsthead 

\multicolumn{9}{c}%
{{\bfseries \tablename\ \thetable{} -- continued from the previous page}} \\

\hline 
 &  \textbf{MC64} & \multicolumn{3}{c |}{\textbf{Without Heuristic}} & \multicolumn{3}{c |}{\textbf{With Heuristic}}  &	\textbf{HWPM} \\ [1.1ex]
     Matrix & \textbf{Weight}	&	$\frac{\textbf{MCM}}{\textbf{MC64}}$ & $\frac{\textbf{HWPM}}{\textbf{MC64}}$ &  
     \multicolumn{1}{c|}{\textbf{Diff}} & 
     $\frac{\textbf{MCM}}{\textbf{MC64}}$ &
     $\frac{\textbf{HWPM}}{\textbf{MC64}}$ &
     \multicolumn{1}{c|}{\textbf{Diff}} &
     \textbf{Improv.}\\[1.1ex]
\hline 
\endhead

\hline \multicolumn{9}{|r|}{{Continued on the next page}} \\ \hline
\endfoot

\hline 
\endlastfoot

\hline	
barrier2-10	&	113769.90	&	32.11\%	&	63.42\%	&	31.32\%	&	98.64\%	&	99.13\%	&	0.48\%	&	35.71\%	\\
barrier2-11	&	113769.90	&	32.11\%	&	63.45\%	&	31.34\%	&	98.64\%	&	99.13\%	&	0.48\%	&	35.68\%	\\
barrier2-12	&	113769.88	&	32.11\%	&	63.46\%	&	31.35\%	&	98.64\%	&	99.13\%	&	0.48\%	&	35.67\%	\\
barrier2-9	&	113769.90	&	32.11\%	&	63.46\%	&	31.36\%	&	98.64\%	&	99.13\%	&	0.48\%	&	35.66\%	\\
laminar\_duct3D	&	62809.55	&	31.43\%	&	57.59\%	&	26.17\%	&	90.01\%	&	92.72\%	&	2.71\%	&	35.12\%	\\
Raj1	&	263739.51	&	47.19\%	&	71.14\%	&	23.95\%	&	100.00\%	&	100.00\%	&	0.00\%	&	28.86\%	\\
ASIC\_100k	&	99340.00	&	29.91\%	&	78.73\%	&	48.83\%	&	100.00\%	&	100.00\%	&	0.00\%	&	21.27\%	\\
g7jac200	&	43808.97	&	71.52\%	&	71.71\%	&	0.19\%	&	89.38\%	&	91.44\%	&	2.06\%	&	19.73\%	\\
g7jac180	&	39575.87	&	71.71\%	&	73.27\%	&	1.56\%	&	89.67\%	&	91.59\%	&	1.92\%	&	18.32\%	\\
g7jac180sc	&	43446.76	&	72.81\%	&	75.96\%	&	3.14\%	&	93.54\%	&	93.60\%	&	0.06\%	&	17.64\%	\\
g7jac200sc	&	48318.70	&	72.84\%	&	75.88\%	&	3.04\%	&	93.36\%	&	93.50\%	&	0.14\%	&	17.62\%	\\
twotone	&	110113.50	&	27.39\%	&	79.91\%	&	52.52\%	&	97.48\%	&	97.48\%	&	0.00\%	&	17.57\%	\\
scircuit	&	170997.65	&	30.66\%	&	83.80\%	&	53.15\%	&	100.00\%	&	100.00\%	&	0.00\%	&	16.20\%	\\
shyy161	&	76388.42	&	38.78\%	&	85.17\%	&	46.39\%	&	100.00\%	&	100.00\%	&	0.00\%	&	14.83\%	\\
pre2	&	639561.59	&	43.55\%	&	86.36\%	&	42.81\%	&	97.51\%	&	99.33\%	&	1.82\%	&	12.97\%	\\
FullChip	&	2986883.38	&	25.53\%	&	87.15\%	&	61.62\%	&	100.00\%	&	100.00\%	&	0.00\%	&	12.85\%	\\
hvdc2	&	189239.78	&	26.63\%	&	88.12\%	&	61.49\%	&	95.58\%	&	99.85\%	&	4.27\%	&	11.73\%	\\
rajat28	&	87110.15	&	36.47\%	&	88.40\%	&	51.93\%	&	99.80\%	&	99.96\%	&	0.17\%	&	11.56\%	\\
rajat25	&	87075.66	&	36.96\%	&	88.43\%	&	51.47\%	&	99.85\%	&	99.96\%	&	0.11\%	&	11.53\%	\\
PR02R	&	151816.48	&	73.00\%	&	73.00\%	&	0.00\%	&	83.93\%	&	84.46\%	&	0.53\%	&	11.46\%	\\
rajat20	&	86809.29	&	37.29\%	&	88.69\%	&	51.40\%	&	99.83\%	&	99.97\%	&	0.14\%	&	11.28\%	\\
LeGresley\_87936	&	87409.32	&	32.25\%	&	90.00\%	&	57.75\%	&	95.73\%	&	99.63\%	&	3.91\%	&	9.63\%	\\
transient	&	178842.78	&	23.05\%	&	92.28\%	&	69.22\%	&	99.99\%	&	99.99\%	&	0.01\%	&	7.72\%	\\
mark3jac140sc	&	53859.12	&	82.54\%	&	85.51\%	&	2.97\%	&	92.92\%	&	93.10\%	&	0.18\%	&	7.58\%	\\
mark3jac120sc	&	46298.52	&	82.62\%	&	85.55\%	&	2.93\%	&	92.89\%	&	93.07\%	&	0.18\%	&	7.52\%	\\
mark3jac120	&	44129.34	&	87.18\%	&	88.53\%	&	1.34\%	&	95.20\%	&	95.23\%	&	0.03\%	&	6.70\%	\\
mark3jac140	&	51388.64	&	87.29\%	&	88.56\%	&	1.27\%	&	94.82\%	&	94.95\%	&	0.13\%	&	6.38\%	\\
Hamrle3	&	1159868.91	&	90.48\%	&	93.98\%	&	3.50\%	&	99.84\%	&	99.84\%	&	0.00\%	&	5.85\%	\\
rajat24	&	358036.22	&	24.79\%	&	95.60\%	&	70.81\%	&	99.96\%	&	99.99\%	&	0.03\%	&	4.39\%	\\
mac\_econ\_fwd500	&	159806.83	&	89.03\%	&	89.07\%	&	0.05\%	&	92.49\%	&	92.88\%	&	0.39\%	&	3.81\%	\\
ASIC\_680k	&	676120.73	&	90.89\%	&	96.60\%	&	5.71\%	&	99.86\%	&	99.90\%	&	0.04\%	&	3.29\%	\\
bayer01	&	50614.17	&	93.06\%	&	94.86\%	&	1.80\%	&	96.12\%	&	97.36\%	&	1.23\%	&	2.49\%	\\
hcircuit	&	105676.00	&	43.68\%	&	97.66\%	&	53.98\%	&	100.00\%	&	100.00\%	&	0.00\%	&	2.34\%	\\
RM07R	&	366956.93	&	88.73\%	&	91.54\%	&	2.80\%	&	92.55\%	&	92.84\%	&	0.29\%	&	1.31\%	\\
ASIC\_320k	&	321821.00	&	60.64\%	&	99.15\%	&	38.51\%	&	100.00\%	&	100.00\%	&	0.00\%	&	0.85\%	\\
barrier2-3	&	111224.20	&	66.04\%	&	98.41\%	&	32.37\%	&	98.61\%	&	99.11\%	&	0.50\%	&	0.70\%	\\
barrier2-1	&	111224.26	&	66.04\%	&	98.41\%	&	32.37\%	&	98.61\%	&	99.11\%	&	0.50\%	&	0.70\%	\\
barrier2-2	&	111224.20	&	66.04\%	&	98.42\%	&	32.37\%	&	98.61\%	&	99.11\%	&	0.50\%	&	0.69\%	\\
lhr71	&	68715.55	&	79.78\%	&	98.33\%	&	18.55\%	&	81.30\%	&	98.98\%	&	17.68\%	&	0.64\%	\\
barrier2-4	&	111239.64	&	66.05\%	&	98.42\%	&	32.37\%	&	98.60\%	&	99.02\%	&	0.42\%	&	0.60\%	\\
Chebyshev4	&	66210.26	&	99.40\%	&	99.69\%	&	0.29\%	&	99.92\%	&	100.00\%	&	0.08\%	&	0.31\%	\\
lhr71c	&	66419.58	&	73.45\%	&	95.65\%	&	22.20\%	&	90.53\%	&	95.90\%	&	5.37\%	&	0.25\%	\\
torso1	&	116105.15	&	99.93\%	&	99.93\%	&	0.00\%	&	100.00\%	&	100.00\%	&	0.00\%	&	0.07\%	\\
ohne2	&	181166.98	&	66.64\%	&	99.87\%	&	33.24\%	&	99.71\%	&	99.91\%	&	0.20\%	&	0.04\%	\\
TSOPF\_RS\_b39\_c30	&	60093.68	&	50.03\%	&	99.97\%	&	49.94\%	&	50.28\%	&	100.00\%	&	49.72\%	&	0.03\%	\\
crashbasis	&	159872.83	&	99.98\%	&	99.98\%	&	0.00\%	&	99.98\%	&	100.00\%	&	0.02\%	&	0.02\%	\\
para-4	&	152889.99	&	66.45\%	&	99.10\%	&	32.66\%	&	98.24\%	&	99.10\%	&	0.86\%	&	0.00\%	\\
para-8	&	155589.51	&	67.01\%	&	99.14\%	&	32.13\%	&	98.28\%	&	99.14\%	&	0.87\%	&	0.00\%	\\
para-9	&	155588.72	&	67.01\%	&	99.14\%	&	32.13\%	&	98.28\%	&	99.14\%	&	0.86\%	&	0.00\%	\\
para-6	&	155585.52	&	67.01\%	&	99.12\%	&	32.11\%	&	98.28\%	&	99.12\%	&	0.84\%	&	0.00\%	\\
para-5	&	155585.52	&	67.01\%	&	99.12\%	&	32.11\%	&	98.28\%	&	99.12\%	&	0.84\%	&	0.00\%	\\
para-7	&	155585.52	&	67.01\%	&	99.12\%	&	32.11\%	&	98.28\%	&	99.12\%	&	0.84\%	&	0.00\%	\\
para-10	&	155582.54	&	67.01\%	&	99.12\%	&	32.11\%	&	98.28\%	&	99.12\%	&	0.84\%	&	0.00\%	\\
largebasis	&	423555.70	&	81.82\%	&	100.00\%	&	18.18\%	&	100.00\%	&	100.00\%	&	0.00\%	&	0.00\%	\\
ibm\_matrix\_2	&	51446.65	&	97.76\%	&	99.97\%	&	2.21\%	&	99.74\%	&	99.97\%	&	0.23\%	&	0.00\%	\\
matrix-new\_3	&	125326.51	&	98.80\%	&	99.99\%	&	1.19\%	&	99.92\%	&	99.99\%	&	0.07\%	&	0.00\%	\\
lung2	&	109430.47	&	99.53\%	&	99.54\%	&	0.01\%	&	99.53\%	&	99.54\%	&	0.01\%	&	0.00\%	\\
webbase-1M	&	995848.08	&	99.97\%	&	99.98\%	&	0.01\%	&	99.97\%	&	99.98\%	&	0.01\%	&	0.00\%	\\
language	&	399130.00	&	99.96\%	&	99.96\%	&	0.00\%	&	99.96\%	&	99.96\%	&	0.00\%	&	0.00\%	\\
Baumann	&	110852.43	&	99.83\%	&	99.83\%	&	0.00\%	&	99.83\%	&	99.83\%	&	0.00\%	&	0.00\%	\\
tmt\_unsym	&	3671299.99	&	25.00\%	&	25.00\%	&	0.00\%	&	25.00\%	&	25.00\%	&	0.00\%	&	0.00\%	\\

\hline
\hline 
\textbf{Average}	&		&	\textbf{73.10\%}	&	\textbf{91.95\%}	&	\textbf{18.85\%}	&	\textbf{96.60\%}	&	\textbf{97.85\%}	&	\textbf{1.24\%}	&	\textbf{5.89\%}	\\
\textbf{Minimum}	&		&	\textbf{18.84\%}	&	\textbf{25.00\%}	&	\textbf{0.00\%}	&	\textbf{25.00\%}	&	\textbf{25.00\%}	&	\textbf{0.00\%}	&	\textbf{0.00\%}	\\
\textbf{Maximum}	&		&	\textbf{100.00\%}	&	\textbf{100.00\%}	&	\textbf{77.31\%}	&	\textbf{100.00\%}	&	\textbf{100.00\%}	&	\textbf{49.72\%}	&	\textbf{35.71\%}	\\
  
\label{table:weights-all}
 \end{longtable}
 \end{center}
 
 \normalsize

When the tie-breaking heuristic is employed, the approximation ratios of MCM improve significantly: on average from 73.10\% without heuristic is improved to 96.60\% with heuristic. 
With heuristic, the performance of HWPM also improves on average by 5.89\% as shown in column 9 of Table~\ref{table:weights-all}. Therefore, the heuristic not only directly benefits MCM, but also indirectly benefits the alternating cycles algorithm by providing superior initial perfect matchings. 
However, column 8 in Table~\ref{table:weights-all} shows that the WIAC algorithm with heuristic made modest improvement relative to improvement shown in column 5.
For example, the average improvement with the heuristic is just 1.24\% (column 8), whereas the average improvement without the heuristic is just 18.85\% (column 5).
This is expected because MCM with heuristic already provides heavy weights leaving the WIAC algorithm with little room for improvement. 
rajat31 displays a striking example of the impact of the heuristic.
For this graph, the approximation ratio improves by 77.31\% without the heuristic, but does not improve at all when the heuristic is used. 
Therefore, for rajat31, the WIAC algorithm just needs one iteration to check if the MCM weight can be improved. 
Similarly, the MCM algorithm with the heuristic does not leave any room for improvement for ASIC\_100k. 
We observed that, for ASIC\_100k and rajat31, the MCM algorithm has many choices for cardinality-increasing augmentations, making these problems more susceptible to the heuristic.

The last column in Table~\ref{table:weights-all} shows the overall weight improvement of the HWPM when the heuristic is used. 
This column shows that the HWPM algorithm returns a superior solution when it is initialized with a relatively heavy-weight perfect matching.  
Overall, the heuristic has positive influence on the performance of the HWPM algorithm. Hence, unless otherwise stated, the heuristic is used in all of our remaining experiments in the paper.

{\bf Approximation ratios for bigger matrices.} Table~\ref{table:approx-ratio} shows the weights of matchings obtained from MC64 and HWPM for bigger matrices from Table~\ref{table:problem-statistics}. 
For 9 out of 16 matrices in Table~\ref{table:problem-statistics}, HWPM finds a matching with the optimum weight. 
Approximate weights of other matrices are also close to the optimum.
Hence, our algorithm can successfully substitute MWPM algorithms in many practical problems without sacrificing the quality of the matchings.
}

\begin{table}[!t]
 \centering
 \caption{The weight of matchings from HWPM and MC64. 
}
 \begin{tabular}{@{} l r r  r  @{}}
    \toprule
    Matrix & MC64	&	HWPM &	Approx. ratio \\

 \toprule

memchip	&	2,707,524 &	2,707,520	&	~100\%  \\
rajat31	&	4,690,002 &	4,690,002	&	100\%  \\
Freescale2	&	2,994,270	&	2,989,080	& 99.98\%	 \\
boneS10	&	914,898 &	914,898	&	100\%\\
Serena		&	1,391,344 & 1,391,340	& 100\% \\
audikw\_1	&	943,624 &	943,624	&	100\% \\
dielFilterV3real	&	1,102,796 &	1,102,796	& 100\%	 \\
Flan\_1565	&	1,564,794 &	1,564,794	&	100\% \\
circuit5M	&	5,557,920 &	5,557,890	&	99.99\%\\

Li7Nmax6	&	663,526 & 663,521	& 99.99\%\\
HV15R	&	1,877,709 & 1,770,960	&	94.31\%\\
perf008cr	&	7,902,327	& 7,902,327 &  100\%\\
nlpkkt240	&	27,762,507 &	27,710,011 & 99.81\%\\
Nm7	&	4,079,730 & 4,079,730 & 100\% \\
NaluR3	&	17,598,889 & 17,598,889 &	100\%\\
A05	& 1,140,660 & 1,140,660	& 100\% \\
      \toprule
  \end{tabular}
\label{table:approx-ratio}
 \end{table}

\subsection{Load balancing by randomly permuting the matrix} 
\label{sec:load-balance}
In our experiments, we balance load across processes by randomly permuting rows and columns of the input matrix  before running the matching algorithms. 
Let $m_{max}$ be the maximum number of nonzeros stored in a process and $m_{avg} = m/p$ be the average number of nonzeros expected in a process under perfect load balance. Then, $\frac{m_{max}}{m_{avg}}$ measures the degree of load imbalance given a distribution of the matrix among the processes. The higher the ratio, the more imbalanced the distribution is. A ratio equal to one denotes perfect load balance. We computed this load-imbalance ratio with 256 processes for all matrices in our test suite. \finalrevision{The arithmetic mean of the load-imbalance ratio is 10 before permutation and 1.2 after random permutation of the matrix. 
Load imbalance is also insensitive to matrix size. 
For example, A05, the biggest matrix we experimented with, has load imbalance of 10.67 before permutation and 1.08 after the random permutation.}
Thus, for the real-world instances in our test set, the random permutation is very effective at distributing a matrix uniformly across processes.  

\subsection{Performance of the parallel HWPM algorithm}
\begin{figure}[!t]
 \begin{center}

    \includegraphics[scale = .2 ]{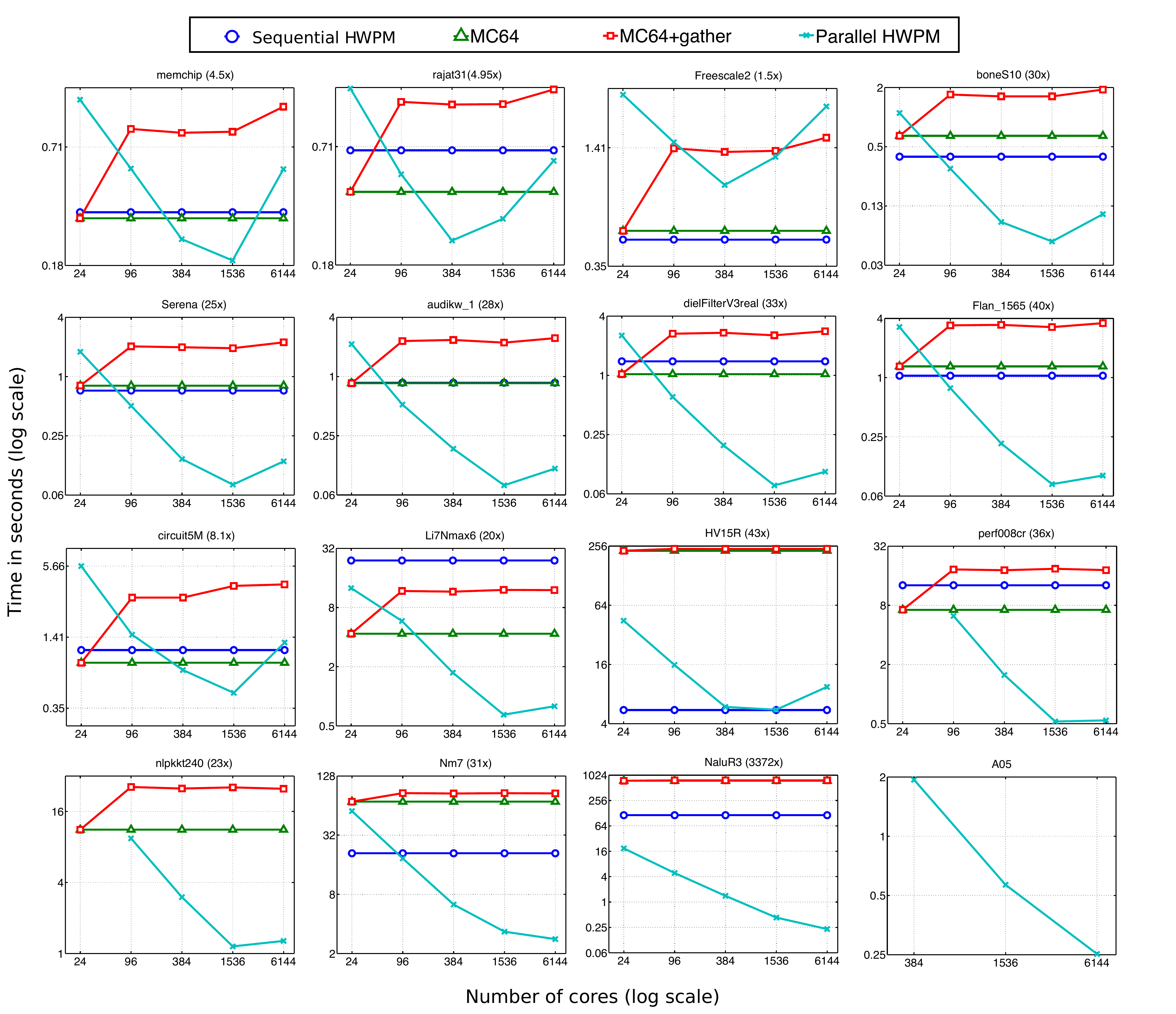}
  \caption{Comparing the parallel HWPM algorithm with sequential HWPM and MC64 on Edison. HWPM time includes the computation of a perfect matching and the newly developed weight increasing algorithm.    A red line plots the time to gather a distributed matrix on a node and then run MC64. \revision{For each matrix, the best speedup attained by the parallel HWPM algorithm relative to MC64+\textsc{gather} time is shown at the top of the corresponding subplot.} Four MPI processes per node and 6 threads per process were used. Matrices are arranged in ascending order by $nnz$ first from left to right and then from top to bottom.}
  \label{fig:compare-algo-edison}
 \end{center}
\end{figure}

Figure~\ref{fig:compare-algo-edison} compares the running time of the parallel HWPM algorithm with MC64 and sequential HWPM for all 16 matrices from Table~\ref{table:problem-statistics}.  Here MC64+\textsc{gather} lines include the time to gather the whole matrix on a single node, representing the total time needed to compute MWPM sequentially in a distributed memory setting. Experiments in Figure~\ref{fig:compare-algo-edison} were run on Edison. 

First, we discuss the performance of matching algorithms on bigger matrices where a distributed memory algorithm is urgently needed.
MC64 failed with A05, the largest matrix in our suite, because of insufficient memory in a single node of Edison. The same is true for sequential HWPM.
For other large matrices, parallel HWPM is significantly faster than its sequential competitors. For some problems, this is true even on a single node. For example, when computing a matching on NaluR3 on a single node of Edison, parallel HWPM is $40\times$ and $6\times$ faster than MC64 and sequential HWPM, respectively. On 256 nodes (6144 cores), our parallel algorithm becomes $3372\times$ and $384\times$ faster than MC64 and sequential HWPM. Note that HWPM is able to find the optimum solution on NaluR3. Hence, in this case the drastic reduction in running time comes for free, without sacrificing any matching quality.
For other large matrices (with $nnz$\footnote{For consistency with established notation, we use $nnz$ (i.e.,~the number of nonzeroes) when referring to matrices. This number is identical to the number of edges (traditionally labeled $m$) in the corresponding graph.} greater than 200M), parallel HWPM runs at least $20\times$ faster than MC64+\textsc{gather} on high concurrency.
This observation also holds for most matrices in the second and third rows in Figure~\ref{fig:compare-algo-edison}.

\begin{figure}[!t]
 \begin{center}
    \includegraphics[scale = .2 ]{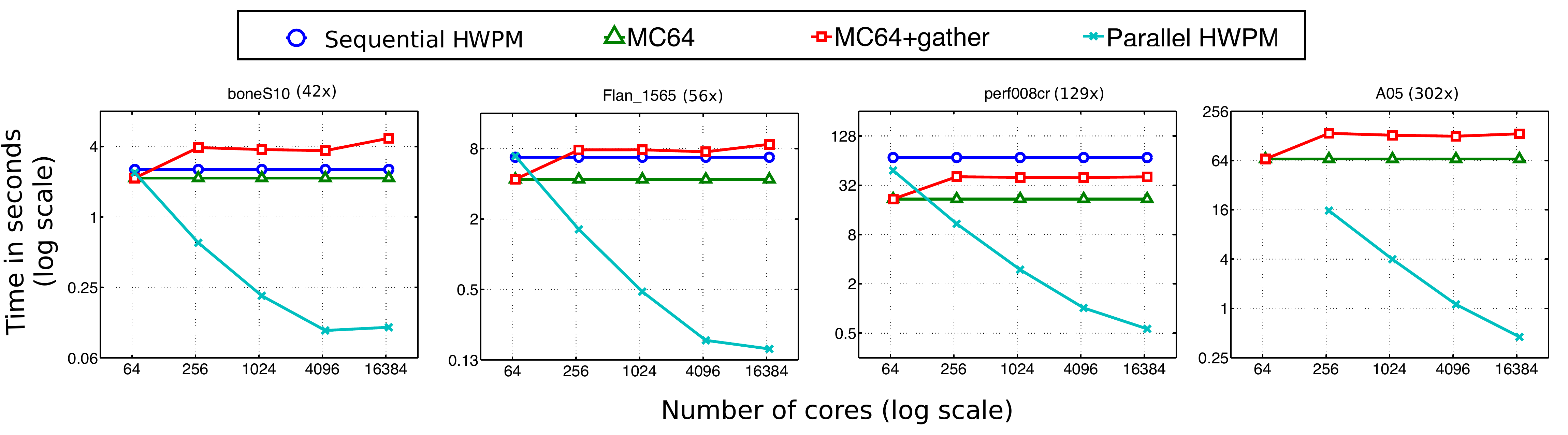} 
  \caption{Comparing parallel HWPM with two sequential algorithms for four representative matrices on Cori-KNL.  The best speedup of the parallel HWPM algorithm relative to MC64+\textsc{gather} time is shown at the top of the corresponding subplot.  See the caption of Figure~\ref{fig:compare-algo-edison} for further detail.}
  \label{fig:compare-algo-Cori}
 \end{center}
\end{figure}

\begin{figure*}[!t]
 \begin{center}
  \subfloat {
    \includegraphics[scale = .4 ]{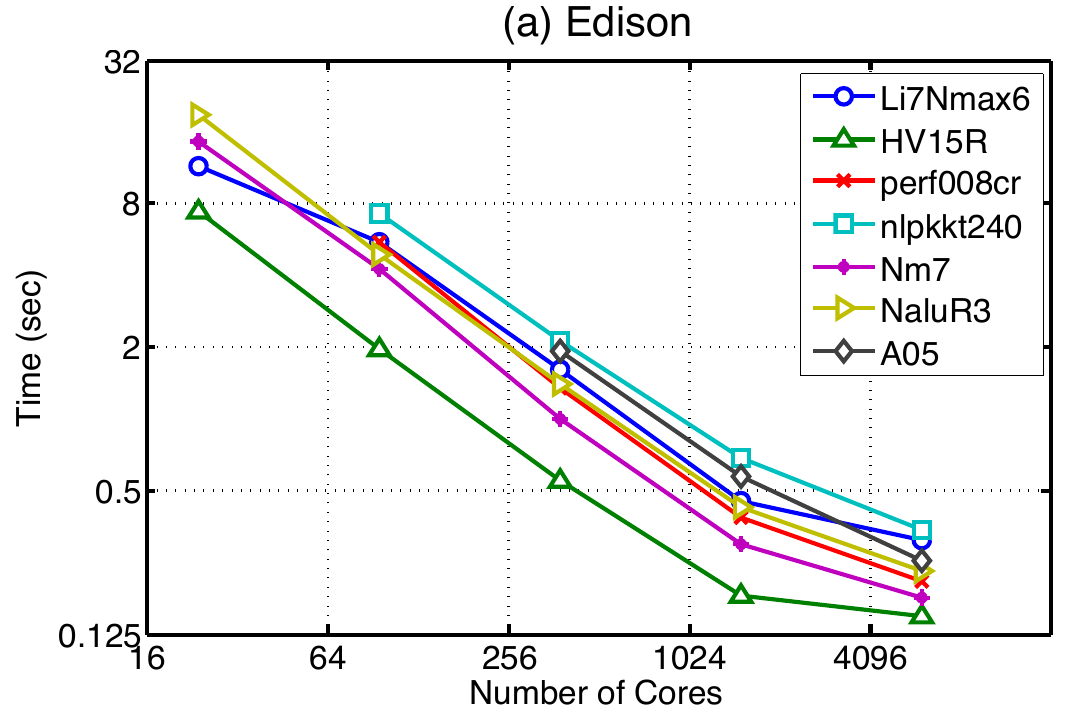}
  } ~
  \subfloat {
    \includegraphics[scale = .4]{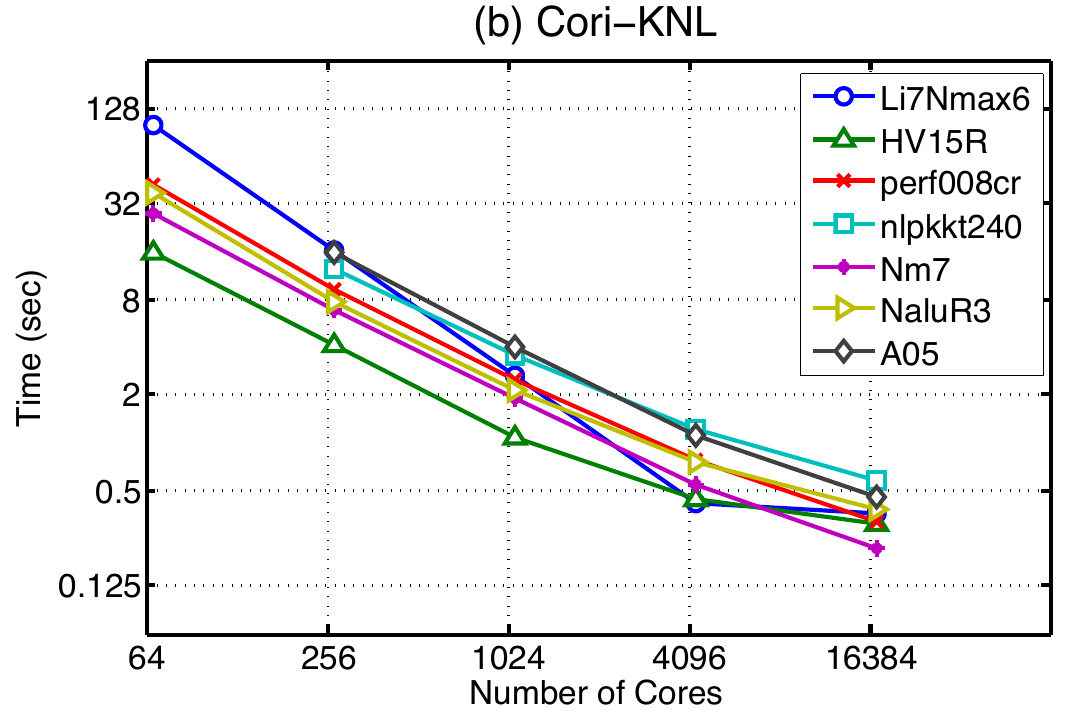}
  }
  \caption{Strong scaling of the weight increasing alternating cycles algorithm (without the perfect matching) for the largest seven matrices from Table~\ref{table:problem-statistics}. The scalability plots starts from one node. Four MPI processes are used in each node of Edison and Cori-KNL.}
  \label{fig:cycleonly-big7-scaling}
 \end{center}
\end{figure*}

\begin{figure*}[!t]
 \begin{center}
  \subfloat {
    \includegraphics[scale = .33 ]{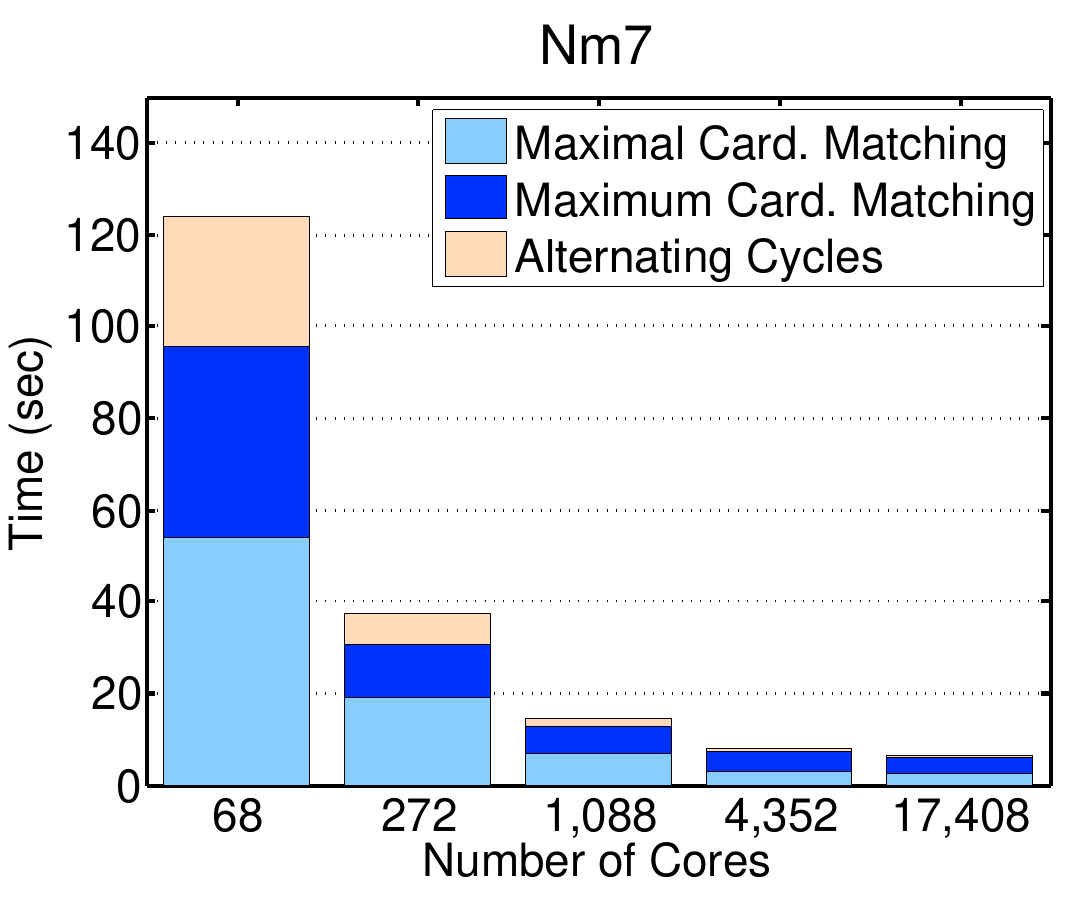}
  } ~
  \subfloat {
    \includegraphics[scale = .33]{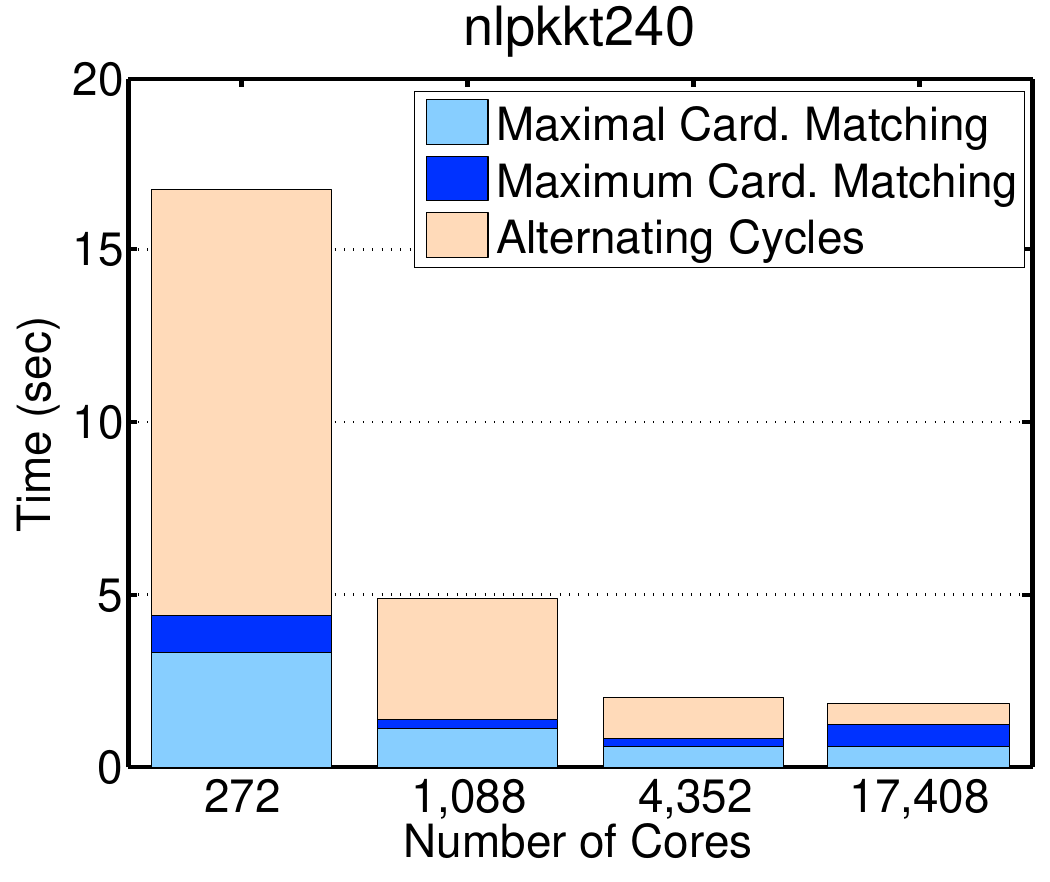}
  }
  \caption{Breakdown of parallel HWPM running time on Cori-KNL.}
  \label{fig:runtime-breakdown}
 \end{center}
\end{figure*}

On smaller matrices (e.g., those in the first row of Figure~\ref{fig:compare-algo-edison}), the performance gain from the parallel algorithm is not as dramatic as with bigger matrices.
This is expected as it is hard for parallel algorithms to reduce a subsecond sequential running time. 
However, for all matrices except Freescale2, parallel HWPM runs faster than MC64+\textsc{gather} on high concurrency. 
Hence our algorithm is competitive on smaller matrices and runs significantly faster on bigger matrices. 

While most instances show excellent scalability, there are two outliers, Freescale2 and HV15R, where parallel HWPM does  not run faster than the sequential HWPM.
For both matrices, parallel HWPM spends more than 80\% of the running time on finding an initial perfect matching using a distributed-memory MCM algorithm~\cite{matchingipdps16}.
Obtaining perfect matchings on these two matrices requires searching for long paths going through many processors, which is hard to parallelize.
Nevertheless, even for these matrices, parallel HWPM remains competitive or significantly faster than MC64+\textsc{gather}, which is the main competitor of our algorithm in practice.

The matching algorithms discussed above show similar performance trends on Cori-KNL, as shown in Figure~\ref{fig:compare-algo-Cori}.  
For example, on the perf00cr matrix, parallel HWPM runs $72\times$ and  $34\times$ faster than MC64+\textsc{gather} on 17,408 and 6,144 cores on Cori-KNL and Edison, respectively. 
On Cori-KNL, MC64 is able to compute a matching for the A05 matrix in 135 seconds, whereas parallel HWPM algorithm took just 0.44 seconds on 256 nodes of Cori-KNL.

\subsection{Scalability of parallel HWPM algorithm}
Unless otherwise noted, we report the speedup of the parallel HWPM algorithm relative to its running time on a single node. HWPM still runs in parallel on a single node using 4 MPI processes and employs multithreading within a process.   

At first, we discuss the scalability of the complete HWPM algorithm (including the time to compute the initial perfect matching) as shown in Figure~\ref{fig:compare-algo-edison} and~\ref{fig:compare-algo-Cori}. 
HWPM achieves a $19\times$ speedup
on average over 13 matrices that were able to run on a single node of Edison.
However, the performance improvement is more significant on bigger matrices. For example, our algorithm 
attains the best speedup of $82\times$ for NaluR3 on 256 nodes on Edison.
On 256 nodes of Cori-KNL, HWPM achieves $30\times$ speedup
on average over 14 matrices that were able to run on a single node. The best speedups on Cori-KNL are attained on Li7Nmax6 ($114\times$) and NaluR3 ($97\times$).
For the A05 matrix, as we go from 16 nodes to 256 nodes, parallel HWPM runs $8\times$ and $9\times$ faster on Edison and Cori-KNL, respectively.

Since the primary contribution of the paper is the weight increasing alternating 4-cycles algorithm, we show its scalability separately in Figure~\ref{fig:cycleonly-big7-scaling}. 
We only show results for matrices with more than 200M nonzeros.
For all matrices in Figure~\ref{fig:cycleonly-big7-scaling}, the parallel weight increasing alternating cycles algorithm attains more than a $50\times$ speedup on both Edison and Cori-KNL. Among these matrices, the best speedup of $260\times$ was observed on 256 nodes of Cori-KNL for the Li7Nmax6 matrix, highlighting the excellent scalability of the newly developed algorithm.

\subsection{Insights on the performance of parallel HWPM algorithm}
In order to study the behavior of parallel HWPM, we break down the execution time of two instances in Figure~\ref{fig:runtime-breakdown}.  
For both instances, the newly developed alternating cycles algorithm scales very well. This trend is also observed for most of the matrices, as was explained before.
However, the maximum cardinality matching algorithm that is used to obtain an initial perfect matching starts to become the bottleneck on high concurrency. 
As studied extensively in a prior work~\cite{matchingipdps16}, the performance of the parallel MCM algorithm suffers if it needs to search long augmenting paths that span many processors. 
 

\subsection{Performance of HWPM as a pre-pivoting tool for a distributed sparse direct solver}
As stated in Section~\ref{sec:intro}, our motivation for HWPM comes
from the need for parallel pre-pivoting of large entries to the main diagonal
to improve stability of sparse direct solvers. We now evaluate how 
HWPM performs relative to MC64 in this regard, using the SuperLU\_DIST direct solver.
We investigate both the accuracy of the solution and the running time of the static-pivoting step when MC64 and HWPM are used with  SuperLU\_DIST.

\begin{table}[!t]
 \centering
 \caption{ A summary of the impact of various matching solutions on the accuracy of the solver with an extended set of matrices considered in Table~\ref{table:weights-all}. The detail result is available in the supplementary file. A matching is used to permute rows of a matrix before factorization. A permutation fails when the relative solution error is close to 1. 
 Rows in this table consider the success and failure with MCM (with heuristic) and HWPM. 
 Columns consider the success and failure with MC64 Option 5 with and without scaling.  
 }

 \begin{tabular}{@{} r | r r  r r@{}}
    \bottomrule
    & \multicolumn{2}{c}{MC64-no-scaling} & \multicolumn{2}{c}{MC64-scaling}
    \\
  
    & Fail   &   Success &         Fail    &    Success \\
    \cline{1-5}
    MCM(Fail) and HWPM(Fail)  &  31    &  4          &        21    &     14 \\

MCM(Fail) and HWPM(Success)  &  4   &  3       &          0     &    7  \\

MCM(Success) and HWPM(Fail)  &   0  &    0        &          0     &     0 \\

MCM(Success) and HWPM(Success)  &  12    &  48       &      0      &    60 \\
 \toprule

  \end{tabular}
\label{table:superlu-perf-all-summary}
 \end{table}

\begin{table}[!t]
 \centering
 \caption{Weights of matchings from MC64 and HWPM when the sum of logarithm of weights is maximized and SuperLU\_DIST relative error of the solution. A relative error of 1
 indicates no significant digit is correct. Note that we do not use the scaling of MC64 Option 5.
 For other matrices in Table~\ref{table:problem-statistics}, we could not get a solution from SuperLU\_DIST. 
 }

 \begin{tabular}{@{} l r r | r r   r   @{}}
    \toprule
    & \multicolumn{2}{c|}{Matching weight } & \multicolumn{3}{c}{Relative error}  \\
    Matrix & MC64	&	HWPM & MC64	&	HWPM &  MCM (with heuristic) \\
 \toprule
memchip	& $0.00$ & $0.00$ & 	$7.68{\times} 10^{-14}$ &	$7.68{\times} 10^{-14}$	& $6.15{\times} 10^{-13}$\\
rajat31	& $0.00$ & $0.00$ &	$5.65{\times} 10^{-11}$  &	$5.66{\times} 10^{-11}$	& $7.18{\times} 10^{-11}$\\
Freescale2	& -11982.80	& -15122.14 &	$1$	&	$1$		& $1$\\
boneS10	& $0.00$ & $0.00$ &	$1.35{\times} 10^{-09}$ &	$1.36{\times} 10^{-09}$ 	& $1.36{\times} 10^{-09}$\\
Serena		& -6.33	& -6.33 &	$2.41{\times} 10^{-13}$	 & $2.96{\times} 10^{-13}$ & $2.96{\times} 10^{-13}$\\
audikw\_1	& -81.90	& -81.90 &		$1.49{\times} 10^{-10}$  &	 $1.25{\times} 10^{-10}$ &  $1.25{\times} 10^{-10}$\\
dielFilterV3real & -28.97 &	-28.97	&	$2.30{\times} 10^{-11}$ &	$2.39{\times} 10^{-11}$	& 	 $2.39{\times} 10^{-11}$\\
Flan\_1565	& $0.00$ & $0.00$ &	$1.93{\times} 10^{-10}$ &	$1.93{\times} 10^{-10}$	& $1.93{\times} 10^{-10}$	\\
circuit5M	& -1298.52 &	-2491.49 &	$5.18{\times}10^{-09}$ &	$1.13{\times} 10^{-08}$	& $1.17{\times} 10^{-08}$\\
perf008cr & $0.00$ & $0.00$	&	$6.56{\times} 10^{-10}$	& $7.09{\times} 10^{-10}$ &  $7.09{\times} 10^{-10}$\\
A05	& -780638	 & -780638 &  $2.85{\times} 10^{-04}$ & $3.62{\times} 10^{-04}$	&  $3.62{\times} 10^{-04}$\\
      \toprule
  \end{tabular}
\label{table:superlu-perf}
 \end{table}

{\bf Accuracy of the solution.}
To compute the accuracy of solutions for a matrix $A$, we set the true
solution vector $x_{true} = [1, \ldots, 1]^T$, then generate the right-hand
side vector by computing $b = A x_{true}$. 
We use the LAPACK-style simple
equilibration scheme to compute the row and column scalings $D_r$ and 
$D_c$ such that in the scaled matrix $D_r A D_c$, the maximum entry in magnitude of each row or column is 1. Hence, the weights of many matched edges are often 1, and the sum of the logarithms is 0.
We then apply the pre-pivoting strategies of MC64 or HWPM
to compute a row permutation vector $P_r$. Here, the maximization criterion
is the sum of the logarithms of the weights, i.e.~the product of the weights.
The sparsity
reordering $P_c$ is then obtained with METIS using the permuted matrix (graph).
Finally, the LU factorization is computed as
$P_c^T (P_r (D_r A D_c)) P_c = LU$, and the solution $x$ is computed based on the
transformed linear system. The relative solution error
$ \|x - x_{true}\|_\infty / \|x\|_\infty$ is reported in 
 Table~\ref{table:superlu-perf} for large-scale matrices from Table~\ref{table:problem-statistics} and in supplementary Table SM1 for an extended set of matrices.

Now, we study the impact of matchings on the accuracy of solvers.
MC64 can also return dual variables that can be used to perform another equilibration.
Hence, we considered MC64 Option 5 with and without scaling (here, scaling means equilibration done with dual variables). 
Since HWPM does not use dual variables, it is equivalent to MC64 without scaling.
Table ~\ref{table:superlu-perf-all} in the Appendix shows the result with 102 matrices from the SuiteSparse Matrix collection, which is summarized in  Table~\ref{table:superlu-perf-all-summary}.
We observe that the solver is sensitive to the pre-pivoting strategies used. 
The error rate varies with different matrices as well as with different matching solutions.

Overall, MC64 with scaling is successful for most matrices in our test suite. For 14 out of 102 matrices, MC64 with scaling succeeds while HWPM fails. However, for these matrices, dual variable scaling is making a significant difference. For example, the solver still fails for 10 of these 14 matrices if MC64 is used without scaling. 
Therefore,  HWPM (with heuristic) is as successful as MC64 without scaling for most matrices in our test suite. 
Finding meaningful dual variables in HWPM remains our future work.
Note that the solver error with HWPM is usually smaller than that of MCM for most problems (Table ~\ref{table:superlu-perf-all}).
HWPM is also successful for 7 matrices where MCM fails.
Hence, increasing weights by weight increasing alternating cycles helps us improve the accuracy of solvers.  



Table~\ref{table:superlu-perf} shows the relative solution error for bigger matrices in our test suite.
For most matrices, the relative error obtained with HWPM is remarkably close
to that of MC64, with the exception of circuit5M. This can be explained 
by the difference in weights found by MC64 and HWPM. However, for most matrices, the HWPM
weights are identical to the MC64 weights. Recall in Table~\ref{table:approx-ratio},
when the sum of
weights is maximized, HWPM achieves 99.99\% optimum weight for this matrix.
When using the permutation obtained from the ``HWPM-sum'' metric
for circuit5M, the computed solution is as accurate as that of MC64.
In the future, we plan to investigate the performance of the HWPM-sum and
HWPM-product metrics in the solution accuracy of SuperLU\_DIST.

\begin{table}[!t]
 \centering
 \caption{Running time for factorizing a matrix by SuperLU\_DIST and pre-pivoting the matrix by MC64 and HWPM on 4 nodes (96 cores) of Edison. MC64 time includes the time to gather the matrix on a single MPI process (i.e., MC64+\textsc{gather} time in Figure~\ref{fig:compare-algo-edison}). A05 running time is not shown because MC64 failed to run on a single node of Edison.
 The remaining five matrices from Table~\ref{table:problem-statistics} are not shown because we could not get a solution from SuperLU\_DIST (MC64 and HWPM were successful as shown in Figure~\ref{fig:compare-algo-edison}).}

 \begin{tabular}{@{} l r r r @{}}
    \toprule
    Matrix & Factorization	&	MC64+\textsc{gather}  & HWPM\\
 \toprule
memchip	&  10.95 & 0.79 & 0.54\\
rajat31	& 24.31 & 1.20 & 0.51\\
Freescale2	& 16.89 & 1.40 & 1.50 \\
boneS10		& 11.21 & 1.22 & 0.29\\
Serena		& 130.93 & 1.56 & 0.50\\
audikw\_1	& 37.17 & 1.67 & 0.51\\
dielFilterV3real & 16.67 & 1.93 & 0.60	 \\
Flan\_1565	& 32.31 &  2.45 & 0.78	\\
circuit5M	& 18.05 & 2.21 & 1.48\\

perf008cr & 2495.8 & 23.20 & 6.23\\
      \toprule
  \end{tabular}
\label{table:superlu-time}
 \end{table}
 
{\bf Running time of the pre-pivoting step.}
Table~\ref{table:superlu-time} shows the time spent on factorizing a matrix by SuperLU\_DIST and pre-pivoting the matrix by MC64 and HWPM on 4 nodes (96 cores) of Edison. As with other experiments, we use 4 MPI processes per node. MC64 time includes the time to gather the matrix on a single MPI process (i.e., MC64+\textsc{gather} time reported in Figure~\ref{fig:compare-algo-edison}). Running time for A05 is not shown because MC64 failed to run on a single node of Edison. For the problems in Table~\ref{table:superlu-perf}, static pivoting with MC64 takes 1\%-10\% of the total solver time on 4 nodes of Edison. As we increase concurrency, factorization time decreases while pivoting time increases. The increasing trends of MC64 running time have already been demonstrated in Figure~\ref{fig:compare-algo-edison} and Figure~\ref{fig:compare-algo-Cori}.
Hence, in accordance to Amdahl's law, MC64 becomes the bottleneck of SuperLU\_DIST on higher concurrency. 
By contrast, the scalability of HWPM reduces the overhead of static pivoting significantly.
For example, when solving perf008cr on 4 nodes of Edison, MC64 and HWPM takes 1\% and 0.3\% of the total running time of SuperLU\_DIST, respectively. 
With more nodes, HWPM overhead quickly disappears, making it an attractive replacement of MC64 when the quality of matching from HWPM is comparable to that of MC64.

\section{Concluding Remarks}\label{sec:conc}
We presented a new distributed-memory parallel algorithm for the {\em heavy-weight perfect matching} problem on bipartite graphs. That is, for most practical problems, our algorithm returns perfect matchings (if they exist) that have high weights. Our motivation comes from distributed-memory sparse direct solvers where an initial permutation of the sparse matrix that places entries with large absolute values on the diagonal is often performed before the factorization, in order to avoid expensive pivoting during runtime. This process is called static pivoting and the permutation is ideally found using a maximum-weight perfect matching. 
However, previous attempts at parallelizing the exact algorithms met with limited success. Since the perfect matching part is a strict requirement of static pivoting, our algorithm only relaxes the weight requirement.

There are two key reasons for the performance of our algorithm. For the initial phase where we find a perfect matching, we use existing optimized distributed-memory cardinality matching algorithms with minor modifications to maximize the weight when tie-breaking. For the iterative phase that improves on the weights while maintaining maximum cardinality, we restrict the algorithm 
to 4-cycles, and thus avoid traversing long augmenting paths, following the Pettie-Sanders approximation algorithm. Hence, the crux of our paper is an efficient method for finding  weight-increasing $4$-cycles on a bipartite graph in distributed memory. 

In terms of LP-duality, unlike the Hungarian algorithm \cite{Kuhn55}, ours is a {\em primal} method because it maintains a feasible solution. It does not compute a solution to the dual problem (i.e.,~feasible potential). Since the dual values can be used to scale matrices in linear solvers \cite{OLSCHOWKA96}, extending our algorithm to provide such scaling factors is a goal for future work. 

We have tested our algorithm on hundreds of large-scale matrices. While in some cases finding the perfect matching in parallel is expensive, the weight improving algorithm always scales well. Our experiments show that the weight of the matchings we obtain is high enough to replace MC64 for pre-pivoting in linear solvers such as SuperLU\_DIST, thereby making such solvers fully scalable. 

\section{Acknowledgements}
This research was supported in part by the Exascale Computing Project (17-SC-20-SC), a collaborative effort of the U.S. Department of Energy Office of Science and the National Nuclear Security Administration.

\bibliographystyle{siamplain}
\bibliography{AWPM}

\begin{thebibliography}{10}

\bibitem{aktulga2014optimizing}
{\sc H.~M. Aktulga, A.~Bulu{\c{c}}, S.~Williams, and C.~Yang}, {\em Optimizing
  sparse matrix-multiple vectors multiplication for nuclear configuration
  interaction calculations}, in Proceedings of the IPDPS, 2014, pp.~1213--1222.

\bibitem{ABMP}
{\sc H.~Alt, N.~Blum, K.~Mehlhorn, and M.~Paul}, {\em Computing a maximum
  cardinality matching in a bipartite graph in time
  ${O}(n^{1.5}\sqrt{m/\log{n}})$}, Information Processing Letters, 37 (1991),
  pp.~237--240.

\bibitem{AzadB16}
{\sc A.~Azad and A.~Bulu{\c{c}}}, {\em A matrix-algebraic formulation of
  distributed-memory maximal cardinality matching algorithms in bipartite
  graphs}, Parallel Computing, 58 (2016), pp.~117--130.

\bibitem{matchingipdps16}
{\sc A.~Azad and A.~Bulu\c{c}}, {\em Distributed-memory algorithms for maximum
  cardinality matching in bipartite graphs}, in Proceedings of the IPDPS, IEEE,
  2016.

\bibitem{tpds16}
{\sc A.~Azad, A.~Bulu\c{c}, and A.~Pothen}, {\em Computing maximum cardinality
  matchings in parallel on bipartite graphs via tree-grafting}, IEEE
  Transactions on Parallel and Distributed Systems (TPDS), 28 (2017),
  pp.~44--59.

\bibitem{azad2012multithreaded}
{\sc A.~Azad, M.~Halappanavar, S.~Rajamanickam, E.~G. Boman, A.~Khan, and
  A.~Pothen}, {\em Multithreaded algorithms for maximum matching in bipartite
  graphs}, in 2012 IEEE 26th International Parallel and Distributed Processing
  Symposium, IEEE, 2012, pp.~860--872.

\bibitem{Bertsekas}
{\sc D.~P. Bertsekas and D.~A. Casta{\~{n}}{\'{o}}n}, {\em A generic auction
  algorithm for the minimum cost network flow problem}, Comp. Opt. and Appl., 2
  (1993), pp.~229--259.

\bibitem{borvstnik2014sparse}
{\sc U.~Bor{\v{s}}tnik, J.~VandeVondele, V.~Weber, and J.~Hutter}, {\em Sparse
  matrix multiplication: The distributed block-compressed sparse row library},
  Parallel Computing, 40 (2014), pp.~47--58.

\bibitem{CombBLAS}
{\sc A.~Bulu{\c{c}} and J.~R. Gilbert}, {\em The {C}ombinatorial {BLAS}:
  Design, implementation, and applications}, International Journal of
  High-Performance Computing Applications {(IJHPCA)}, 25 (2011).

\bibitem{bulucc2012parallel}
{\sc A.~Bulu{\c{c}} and J.~R. Gilbert}, {\em Parallel sparse matrix-matrix
  multiplication and indexing: Implementation and experiments}, SIAM Journal on
  Scientific Computing, 34 (2012), pp.~C170--C191.

\bibitem{combblas_web}
{\em {The Combinatorial BLAS Library}}.
\newblock \url{https://people.eecs.berkeley.edu/~aydin/CombBLAS/html/}.

\bibitem{ufget}
{\sc T.~A. Davis and Y.~Hu}, {\em The {University} of {Florida} sparse matrix
  collection}, ACM Transactions on Mathematical Software (TOMS), 38 (2011),
  p.~1.

\bibitem{DuanPettieSu2018}
{\sc R.~Duan, S.~Pettie, and H.~Su}, {\em Scaling algorithms for weighted
  matching in general graphs}, {ACM} Transactions on Algorithms, 14 (2018),
  pp.~8:1--8:35.

\bibitem{DuffKayaUcar}
{\sc I.~S. Duff, K.~Kaya, and B.~U{\c{c}}ar}, {\em Design, implementation, and
  analysis of maximum transversal algorithms}, {ACM} Transaction on
  Mathematical Software, 38 (2011), pp.~13:1-- 13:31.

\bibitem{duff2011design}
{\sc I.~S. Duff, K.~Kaya, and B.~U{\c{c}}car}, {\em Design, implementation, and
  analysis of maximum transversal algorithms}, ACM Transactions on Mathematical
  Software (TOMS), 38 (2011), p.~13.

\bibitem{MC64}
{\sc I.~S. Duff and J.~Koster}, {\em On algorithms for permuting large entries
  to the diagonal of a sparse matrix}, SIAM Journal on Matrix Analysis and
  Applications, 22 (2001), pp.~973--996.

\bibitem{Goldberg}
{\sc A.~V. Goldberg and R.~E. Tarjan}, {\em A new approach to the maximum flow
  problem}, Journal of the Association for Computing Machinery, 35 (1988),
  pp.~921--940.

\bibitem{HFVTP2012}
{\sc M.~Halappanavar, J.~Feo, O.~Villa, A.~Tumeo, and A.~Pothen}, {\em
  Approximate weighted matching on emerging manycore and multithreaded
  architectures}, The International Journal of High Performance Computing
  Applications, 26 (2012), pp.~413--430.

\bibitem{HoggScott2015}
{\sc J.~Hogg and J.~Scott}, {\em On the use of suboptimal matchings for scaling
  and ordering sparse symmetric matrices}, Numerical Linear Algebra with
  Applications, 22 (2015), pp.~648--663.

\bibitem{Hopcroft}
{\sc J.~E. Hopcroft and R.~M. Karp}, {\em An $n^{\mbox{5/2}}$ algorithm for
  maximum matchings in bipartite graphs}, SIAM Journal on Computing, 2 (1973),
  pp.~225--231.

\bibitem{KLMU2012}
{\sc K.~Kaya, J.~Langguth, F.~Manne, and B.~U\c{c}ar}, {\em Push-relabel based
  algorithms for maximum transversal problem}, Computers \& Operations
  Research, 40 (2013), pp.~1266--1275.

\bibitem{kepner2011graph}
{\sc J.~Kepner and J.~Gilbert}, {\em Graph algorithms in the language of linear
  algebra}, SIAM, 2011.

\bibitem{Kuhn55}
{\sc H.~W. Kuhn}, {\em The hungarian method for the assignment problem}, Naval
  Res. Logistics Quarterly, 2 (1955), pp.~83--97.

\bibitem{LAHM14}
{\sc J.~Langguth, A.~Azad, M.~Halappanavar, and F.~Manne}, {\em On parallel
  push-relabel based algorithms for bipartite maximum matching}, Parallel
  Computing, 40 (2014), pp.~289--308.

\bibitem{LMS2010}
{\sc J.~Langguth, F.~Manne, and P.~Sanders}, {\em Heuristic initialization for
  bipartite matching problems}, ACM Journal of Experimental Algorithmics, 15
  (February, 2010), pp.~1.1--1.22.

\bibitem{LPM2010PC}
{\sc J.~Langguth, M.~M.~A. Patwary, and F.~Manne}, {\em Parallel algorithms for
  bipartite matching problems on distributed memory computers}, Parallel
  Computing, 37 (2011), pp.~820--845.

\bibitem{li2003superlu_dist}
{\sc X.~S. Li and J.~W. Demmel}, {\em {SuperLU}\_{DIST}: A scalable
  distributed-memory sparse direct solver for unsymmetric linear systems}, ACM
  Transactions on Mathematical Software (TOMS), 29 (2003), pp.~110--140.

\bibitem{lin2014towards}
{\sc P.~Lin, M.~Bettencourt, S.~Domino, T.~Fisher, M.~Hoemmen, J.~Hu,
  E.~Phipps, A.~Prokopenko, S.~Rajamanickam, C.~Siefert, et~al.}, {\em Towards
  extreme-scale simulations for low mach fluids with second-generation
  {Trilinos}}, Parallel Processing Letters, 24 (2014), p.~1442005.

\bibitem{pointer}
{\sc F.~Manne and R.~H. Bisseling}, {\em A parallel approximation algorithm for
  the weighted maximum matching problem}, in International Conference on
  Parallel Processing and Applied Mathematics, 2007, pp.~708--717.

\bibitem{suitor}
{\sc F.~Manne and M.~Halappanavar}, {\em New effective multithreaded matching
  algorithms}, in Proceedings of the IPDPS, 2014, pp.~519--528.

\bibitem{MicaliVazirani1980}
{\sc S.~Micali and V.~V. Vazirani}, {\em An {$O(\sqrt{|V|}\cdot|E|)$} algorithm
  for finding maximum matching in general graphs}, in Proceedings of the 21st
  Annual Symposium on Foundations of Computer Science, SFCS '80, Washington,
  DC, USA, 1980, IEEE Computer Society, pp.~17--27.

\bibitem{mitzenmacher2017probability}
{\sc M.~Mitzenmacher and E.~Upfal}, {\em Probability and computing:
  Randomization and probabilistic techniques in algorithms and data analysis},
  Cambridge university press, 2017.

\bibitem{ogielski1993sparse}
{\sc A.~T. Ogielski and W.~Aiello}, {\em Sparse matrix computations on parallel
  processor arrays}, SIAM Journal on Scientific Computing, 14 (1993),
  pp.~519--530.

\bibitem{OLSCHOWKA96}
{\sc M.~Olschowka and A.~Neumaier}, {\em A new pivoting strategy for gaussian
  elimination}, Linear Algebra and its Applications, 240 (1996), pp.~131 --
  151.

\bibitem{pettieSanders}
{\sc S.~Pettie and P.~Sanders}, {\em A simpler linear time 2/3-epsilon
  approximation for maximum weight matching}, Inf. Process. Lett., 91 (2004),
  pp.~271--276.

\bibitem{PF}
{\sc A.~Pothen and C.-J. Fan}, {\em Computing the block triangular form of a
  sparse matrix}, ACM Transactions on Mathematical Software, 16 (1990),
  pp.~303--324.

\bibitem{riedy2010}
{\sc E.~J. Riedy}, {\em Making static pivoting scalable and dependable}, PhD
  thesis, University of California, Berkeley, 2010.

\bibitem{superlu3D}
{\sc P.~Sao, X.~Li, and R.~Vuduc}, {\em A communication-avoiding {3D LU}
  factorization algorithm for sparse matrices}, in Proceedings of the IPDPS,
  IEEE, 2018, pp.~908--919.

\bibitem{sathe2012auction}
{\sc M.~Sathe, O.~Schenk, and H.~Burkhart}, {\em An auction-based weighted
  matching implementation on massively parallel architectures}, Parallel
  Computing, 38 (2012), pp.~595--614.

\bibitem{schrijver2003combinatorial}
{\sc A.~Schrijver}, {\em Combinatorial optimization: Polyhedra and efficiency},
  vol.~24, Springer Science \& Business Media, 2003.

\bibitem{strumpack_web}
{\em {STRUMPACK}: Structured matrices packages}.
\newblock \url{http://portal.nersc.gov/project/sparse/strumpack/}.

\bibitem{superlu_web}
{\em {SuperLU}: Sparse direct solver}.
\newblock \url{http://crd.lbl.gov/~xiaoye/SuperLU/}.

\end{thebibliography}

\newpage
\renewcommand{\thetable}{\Alph{section}\arabic{table}}
\setcounter{table}{0}
\appendix\section{Accuracy of the solution when using matchings from MC64 and HWPM as pre-pivoting tools for SuperLU\_dist}

In the paper, we discussed the impact of exact and approximate matchings on the accuracy of solvers with several large-scale matrices. 
We also presented a summary of 102 small-scale matrices from the SuiteSparse matrix collection in Table 5 of the paper. 
Here, we present the detailed accuracy of those 102 matrices in Table~\ref{table:superlu-perf-all}. We consider MCM (with heuristic), HWPM, MC64 Option 5 with and without scaling (here, scaling means equilibration done with dual variables).
Since HWPM does not use dual variables, it is equivalent to MC64 without scaling.

\begin{center}
\begin{longtable}[H]{@{} l r  r r r @{}}
 \caption{Relative errors when solving an extended set of matrices with SuperLU\_DIST. An MCM, HWPM or MC64 matching is used to permute rows of a matrix before factorization. MCM with the heuristic is used.  A permutation fails when the the relative solution error is close to 1. 
MC64 Option 5 is used both with and without scaling.
 }\\

\hline 
  
  Matrix	&	MCM	&	HWPM	&	MC64	&	MC64	\\
	&		&		&	(no scaling)	&	(scaling)	\\
	

\endfirsthead 

\multicolumn{5}{c}%
{{\bfseries \tablename\ \thetable{} -- continued from the previous page}} \\

\hline 
 Matrix	&	MCM	&	HWPM	&	MC64	&	MC64	\\
	&		&		&	(no scaling)	&	(scaling)	\\
\hline 
\endhead

\hline \multicolumn{5}{|r|}{{Continued on the next page}} \\ \hline
\endfoot

\hline 
\endlastfoot

\hline	
\multicolumn{5}{c }{\bf (Group 1) Success: None; \ \ Fail: All} \\
\hline
torso1	&	4.61E+00	&	1.99E+00	&	2.81E+01	&	3.33E+00\\
thermomech\_dK	&	1.02E+00	&	1.01E+00	&	9.98E-01	&	1.01E+00\\
Chebyshev4	&	1.00E+00	&	1.00E+00	&	1.00E+00	&	1.00E+00\\
Freescale2	&	1.00E+00	&	1.00E+00	&	1.00E+00	&	1.00E+00\\
FullChip	&	1.00E+00	&	1.00E+00	&	1.00E+00	&	1.00E+00\\
laminar\_duct3D	&	1.00E+00	&	1.00E+00	&	1.00E+00	&	1.00E+00\\
mac\_econ\_fwd500	&	1.00E+00	&	1.00E+00	&	1.00E+00	&	1.00E+00\\
mark3jac120	&	1.00E+00	&	1.00E+00	&	1.00E+00	&	1.00E+00\\
mark3jac120sc	&	1.00E+00	&	1.00E+00	&	1.00E+00	&	1.00E+00\\
mark3jac140	&	1.00E+00	&	1.00E+00	&	1.00E+00	&	1.00E+00\\
PR02R	&	1.00E+00	&	1.00E+00	&	1.00E+00	&	1.00E+00\\
pre2	&	1.00E+00	&	1.00E+00	&	1.00E+00	&	1.00E+00\\
rajat20	&	1.00E+00	&	1.00E+00	&	1.00E+00	&	1.00E+00\\
rajat25	&	1.00E+00	&	1.00E+00	&	1.00E+00	&	1.00E+00\\
RM07R	&	1.00E+00	&	1.00E+00	&	1.00E+00	&	1.00E+00\\
shyy161	&	1.00E+00	&	1.00E+00	&	1.00E+00	&	1.00E+00\\
webbase-1M	&	9.97E-01	&	9.97E-01	&	9.97E-01	&	9.97E-01 \\
lhr71	&	9.66E-01	&	9.66E-01	&	9.66E-01	&	9.66E-01 \\
Hamrle3	&	7.50E-01	&	7.50E-01	&	7.50E-01	&	7.50E-01 \\
lhr71c	&	9.66E-01	&	9.66E-01	&	9.66E-01	&	4.14E-01 \\
barrier2-3	&	1.00E+00	&	1.00E+00	&	9.08E-02	&	1.18E-01 \\

\hline
\multicolumn{5}{c}{\bf (Group 2) Success: MC64-scaling;  \ \  Fail: MCM and HWPM}\\
\multicolumn{5}{c}{\bf MC64 without scaling fails in some cases}\\
\hline

g7jac200	&	1.00E+00	&	1.00E+00	&	1.00E+00	&	1.54E-03 \\
barrier2-4	&	1.00E+00	&	1.00E+00	&	6.39E-04	&	7.67E-04 \\
g7jac180	&	1.00E+00	&	1.00E+00	&	1.00E+00	&	6.17E-04 \\
bayer01	&	1.00E+00	&	1.00E+00	&	1.00E+00	&	3.57E-06\\
g7jac200sc	&	1.00E+00	&	1.00E+00	&	1.00E+00	&	2.88E-07\\
g7jac180sc	&	1.00E+00	&	1.00E+00	&	1.00E+00	&	1.89E-07\\
mark3jac140sc	&	1.00E+00	&	1.00E+00	&	1.00E+00	&	8.55E-08\\
rajat17	&	1.00E+00	&	1.00E+00	&	1.00E+00	&	5.08E-08\\
rajat18	&	1.00E+00	&	1.00E+00	&	1.00E+00	&	7.15E-09\\
rajat16	&	1.00E+00	&	1.00E+00	&	1.00E+00	&	4.65E-09\\
twotone	&	1.00E+00	&	1.00E+00	&	1.67E-11	&	1.24E-11\\
hvdc2	&	1.00E+00	&	1.00E+00	&	5.31E-12	&	1.04E-11\\
LeGresley\_87936	&	9.99E-01	&	9.99E-01	&	4.28E-12	&	3.01E-11\\
matrix\_9	&	1.00E+00	&	1.00E+00	&	2.00E+00	&	4.91E-12\\

\hline
\multicolumn{5}{c}{\bf (Group 3) Success: MC64-scaling and HWPM;  \ \  Fail: MCM} \\
\multicolumn{5}{c}{\bf MC64 without scaling fails in some cases}\\
\hline

barrier2-2	&	1.00E+00	&	2.48E-03	&	1.78E-03	&	2.34E-03 \\
barrier2-1	&	1.00E+00	&	2.32E-03	&	1.59E-03	&	2.23E-03 \\
barrier2-12	&	1.00E+00	&	1.95E-04	&	1.00E+00	&	7.33E-04\\
ohne2	&	1.00E+00	&	1.54E-04	&	1.50E-04	&	1.55E-04\\
barrier2-9	&	1.00E+00	&	4.10E-05	&	1.00E+00	&	6.56E-05\\
barrier2-11	&	1.00E+00	&	3.61E-05	&	1.00E+00	&	4.38E-05\\
barrier2-10	&	1.00E+00	&	1.06E-04	&	1.00E+00	&	1.96E-05\\

\hline
\multicolumn{5}{c}{\bf (Group 4) Success: MC64-scaling, HWPM, MCM } \\
\multicolumn{5}{c}{\bf MC64 without scaling fails in some cases}\\
\hline

para-7	&	1.29E-04	&	7.45E-05	&	1.00E+00	&	5.88E-05 \\
ASIC\_680ks	&	1.10E-05	&	1.10E-05	&	1.00E+00	&	1.17E-05 \\
ASIC\_680k	&	6.05E-06	&	9.09E-06	&	7.41E-06	&	1.07E-05 \\
lung2	&	2.35E-06	&	2.67E-06	&	2.20E-06	&	2.07E-06 \\
rajat28	&	1.39E-07	&	1.38E-07	&	1.00E+00	&	1.25E-07 \\
dc3	&	5.09E-08	&	5.08E-08	&	2.92E-08	&	5.20E-08\\
dc1	&	4.65E-08	&	4.87E-08	&	4.91E-08	&	4.86E-08\\
para-8	&	3.26E-04	&	5.84E-04	&	1.00E+00	&	4.47E-04 \\
para-9	&	1.52E-04	&	6.34E-05	&	1.00E+00	&	2.13E-04 \\
para-5	&	1.17E-04	&	9.09E-05	&	1.00E+00	&	1.90E-04	\\
para-6	&	9.53E-05	&	9.44E-05	&	1.00E+00	&	9.25E-05	\\
para-10	&	2.30E-04	&	1.57E-04	&	1.00E+00	&	6.64E-05	\\
dc2	&	4.31E-08	&	4.30E-08	&	2.91E-08	&	4.32E-08	\\
transient	&	1.85E-08	&	2.16E-08	&	2.30E-08	&	1.79E-08	\\
tmt\_unsym	&	1.05E-10	&	3.57E-11	&	5.27E-11	&	1.41E-08	\\
trans5	&	8.57E-09	&	8.58E-09	&	5.94E-09	&	8.67E-09	\\
circuit5M	&	1.13E-08	&	9.99E-01	&	4.40E-09	&	8.21E-09	\\
Freescale1	&	3.18E-09	&	2.11E-09	&	2.23E-09	&	3.15E-09	\\
scircuit	&	2.58E-09	&	3.29E-09	&	3.21E-09	&	2.85E-09	\\
rajat24	&	6.37E-09	&	2.71E-09	&	1.00E+00	&	2.53E-09	\\
rajat23	&	3.62E-09	&	2.96E-09	&	1.00E+00	&	2.50E-09	\\
trans4	&	1.58E-09	&	1.58E-09	&	1.07E-09	&	1.57E-09	\\
ASIC\_320k	&	7.72E-10	&	7.72E-10	&	4.48E-10	&	7.72E-10	\\
rajat26	&	5.77E-11	&	5.19E-11	&	1.00E+00	&	7.12E-10	\\
ASIC\_320ks	&	5.88E-10	&	5.88E-10	&	6.60E-10	&	6.18E-10	\\
bcircuit	&	2.54E-10	&	4.14E-10	&	2.15E-10	&	3.74E-10	\\
Baumann	&	4.95E-10	&	1.01E-10	&	4.02E-10	&	3.46E-10	\\
ASIC\_100ks	&	2.27E-10	&	2.27E-10	&	2.18E-10	&	2.49E-10	\\
Raj1	&	1.31E-10	&	1.39E-10	&	1.02E-10	&	1.64E-10	\\
ASIC\_100k	&	1.91E-10	&	1.91E-10	&	1.28E-10	&	7.16E-11	\\
ecl32	&	6.44E-11	&	7.10E-11	&	1.47E-12	&	6.94E-11	\\
matrix-new\_3	&	5.58E-11	&	5.61E-11	&	4.11E-12	&	4.98E-11	\\
rajat31	&	4.25E-11	&	4.25E-11	&	1.27E-11	&	4.13E-11	\\
language	&	2.05E-11	&	2.04E-11	&	2.01E-11	&	1.86E-11	\\
water\_tank	&	1.06E-11	&	1.29E-11	&	1.60E-11	&	1.47E-11	\\
largebasis	&	1.00E-11	&	9.81E-12	&	7.41E-12	&	9.83E-12	\\
TSOPF\_RS\_b39\_c30	&	3.92E-12	&	3.68E-12	&	4.51E-12	&	3.33E-12	\\
ML\_Geer	&	1.31E-12	&	1.31E-12	&	3.91E-14	&	1.31E-12	\\
epb3	&	1.17E-12	&	1.19E-12	&	2.03E-13	&	1.20E-12	\\
xenon2	&	7.09E-13	&	8.17E-13	&	3.95E-13	&	7.09E-13	\\
ibm\_matrix\_2	&	6.62E-13	&	5.78E-13	&	1.00E+00	&	6.54E-13	\\
ML\_Laplace	&	3.82E-13	&	3.73E-13	&	3.53E-14	&	3.75E-13	\\
hcircuit	&	3.10E-13	&	3.02E-13	&	1.91E-13	&	3.02E-13	\\
atmosmodd	&	1.49E-13	&	1.49E-13	&	1.49E-13	&	2.24E-13	\\
atmosmodj	&	2.01E-13	&	2.02E-13	&	2.02E-13	&	1.81E-13	\\
memchip	&	6.33E-14	&	6.06E-14	&	9.35E-14	&	7.22E-14	\\
venkat50	&	4.13E-14	&	4.71E-14	&	3.62E-14	&	4.97E-14	\\
atmosmodl	&	1.18E-14	&	1.18E-14	&	1.09E-14	&	2.95E-14	\\
venkat25	&	1.93E-14	&	2.24E-14	&	2.16E-14	&	2.64E-14	\\
poisson3Db	&	1.02E-14	&	9.77E-15	&	1.01E-14	&	1.71E-14	\\
circuit5M\_dc	&	1.01E-14	&	9.88E-15	&	1.13E-14	&	1.01E-14	\\
crashbasis	&	8.22E-15	&	9.44E-15	&	8.22E-15	&	8.33E-15	\\
venkat01	&	4.66E-15	&	4.22E-15	&	4.66E-15	&	4.44E-15	\\
torso3	&	3.66E-15	&	3.11E-15	&	2.66E-15	&	4.33E-15	\\
stomach	&	2.00E-15	&	2.00E-15	&	2.00E-15	&	2.22E-15	\\
cage13	&	2.00E-15	&	2.00E-15	&	2.00E-15	&	2.00E-15	\\
FEM\_3D\_thermal2	&	1.78E-15	&	1.78E-15	&	1.55E-15	&	1.78E-15	\\
cage12	&	1.11E-15	&	1.11E-15	&	1.11E-15	&	1.33E-15	\\
torso2	&	1.11E-15	&	8.88E-16	&	8.88E-16	&	1.11E-15

\label{table:superlu-perf-all}
 \end{longtable}
 \end{center}

\end{document}